\documentclass[amsmath,amssymb,showpacs, 12pt]{revtex4}

\usepackage{graphicx}% Include figure files
\usepackage{dcolumn}% Align table columns on decimal point
\usepackage{bm}% bold math
\usepackage{amsmath}
\usepackage{epsf}
\usepackage{epsfig}
\usepackage{amssymb}
\usepackage{color}

\begin{document}

%%%%%%%%%%%%%%%%%%%%%%%%%%%%%%%%%%%
%
%
% Front Matter Section
%
%
%%%%%%%%%%%%%%%%%%%%%%%%%%%%%%%%%%%

\title{Photodetachment of H$^{-}$ by a Short Laser Pulse in Crossed Static Electric and
Magnetic Fields}

\author{Liang-You Peng, Qiaoling Wang,\footnotemark
\footnotetext{Current address: Computer Sciences Corp., c/o Sun
Microsystems Inc., 500 Eldorado Blvd., MS UBRM02-319, Broomfield CO
80021.} and Anthony F. Starace}
%\author{2}
%\author{3}

%\email{Second.Author@institution.edu}

\affiliation{Department of Physics and Astronomy,
    The University of Nebraska-Lincoln, Nebraska 68588-0111, USA}

\date{April 2006}

\begin{abstract}
\begin{center}
{\normalsize \bf{(Submitted to Physical Review A)}}
\end{center}

We present a detailed quantum mechanical treatment of the
photodetachment of H$^{-}$  by a short laser pulse in the presence
of crossed static electric and magnetic fields.  An exact analytic
formula is presented for the final state electron wave function
(describing an electron in both static electric and magnetic fields
and a short laser pulse of arbitrary intensity). In the limit of a
weak laser pulse, final state electron wave packet motion is
examined and related to the closed classical electron orbits in
crossed static fields predicted by Peters and Delos [Phys. Rev. A
\textbf{47}, 3020 (1993)].  Owing to these closed orbit
trajectories, we show that the detachment probability can be
modulated, depending on the time delay between two laser pulses and
their relative phase, thereby providing a means to partially control
the photodetachment process.  In the limit  of a long, weak pulse
(i.e., a monochromatic radiation field) our results reduce to those
of others; however, for this case we analyze the photodetachment
cross section numerically over a much larger range of electron
kinetic energy (i.e., up to 500 cm$^{-1}$) than in previous studies
and relate the detailed structures  both analytically and
numerically to the above-mentioned, closed classical periodic
orbits.  %%
\end{abstract}

% insert suggested PACS numbers in braces on next line

\maketitle

%%%%%%%%%%%%%%%%%%%%%%%%%%%%%%%%%%%
%
%
% Introduction Section
%
%
%%%%%%%%%%%%%%%%%%%%%%%%%%%%%%%%%%%

\section{\label{sec:level1}Introduction}

High resolution studies of atomic Rydberg states in the presence of
external static electric and magnetic fields have proved to be
exceedingly fruitful for the investigation of atomic dynamics
because, owing to the large radial extent and weak binding of atomic
Rydberg levels, the effects of external static fields are much more
significant for Rydberg levels than for atomic ground or low-lying
excited states~\cite{Braun1993}.  Consequently for more than a
quarter century (up to the present) experimentalists and theorists
have been investigating atomic Rydberg spectra in external fields,
including in particular the interesting case of crossed static
electric and magnetic fields~\cite{Braun1993, Crosswhite1979,
Gay1979, Korevaar1983, Clark83, Clark1985, Nessmann1987, Fauth1987,
Hare1988, Wieb89,Gay1989, Vincke1992, Yeaz93, Dippel1994, Uzer1994,
Uzer1996, Connerade1997, Neum97, Uzer1997a, Uzer1997b, Taylor1997,
Sadovskii1998, Wunner1998, Cederbaum1999, Cushman1999, Cushman2000,
Delos2001, Uzer2001, Schmelcher2001, Taylor2002a, Freu02,
Taylor2002b, Valent2003, Wunner2003,
 Abdu04,Conn05}. These latter investigations for the crossed field case include studies
  of motional Stark effects on Rydberg atom spectra in a magnetic
  field~\cite{Crosswhite1979}, of novel, highly excited resonance
  states~\cite{Gay1979, Clark1985, Nessmann1987, Fauth1987, Wieb89, Vincke1992,
  Dippel1994, Connerade1997, Cederbaum1999}, of circular Rydberg states~\cite{Hare1988},
   of Rydberg wave packets in crossed fields~\cite{Gay1989, Yeaz93},
   of non-hydrogenic signatures in Rydberg spectra~\cite{Taylor1997, Wunner1998},
   of doubly-excited states in crossed fields~\cite{Schmelcher2001},
   of recurrence spectra~\cite{Taylor2002a,Freu02, Taylor2002b}, and
   of various aspects of electron dynamics in combined Coulomb and
   crossed static electric and magnetic
   fields~\cite{Braun1993, Uzer1994, Uzer1996, Neum97, Uzer1997a, Uzer1997b,
   Sadovskii1998,Cushman1999, Cushman2000, Delos2001, Uzer2001, Valent2003,
   Wunner2003,Abdu04,Conn05}.

The related problem of photodetachment of a weakly bound electron
(e.g., as in photodetachment of a negative ion) in the presence of
crossed static electric and magnetic fields has been the subject of
fewer investigations despite its having a comparably rich spectrum.
(Note that the weakly bound electron in a negative ion can simply
decay, or become detached, solely due to the presence of the
external static electric and magnetic fields, a process that has
long been studied theoretically, as in,
e.g.,~\cite{Drukarev1972,Popov1998}.) Experimentally, crossed field
effects have been found to be significant in photodetachment of
negative ions in the presence of a static magnetic field owing to
the influence of the motional electric field experienced by the
detached electron~\cite{Blumberg1979, Yuki97}.  The photodetachment
spectrum of H$^-$ in the presence of crossed  static electric and
magnetic fields has been treated theoretically by
Fabrikant~\cite{Fabr91} and by Peters and
Delos~\cite{Peter93,Peter93b}; a generalization to the case of
photodetachment of  H$^-$ in the presence of static electric and
magnetic fields of arbitrary orientation has been given by Liu et
al.~\cite{Liu96,Liu97,Liu97a}.   In each of these works the static
fields are assumed to be sufficiently weak that they do not affect
the relatively compact initial state.  Fabrikant~\cite{Fabr91} gave
the first quantum treatment of single photon detachment in crossed
static electric and magnetic fields using the zero-range potential
model to describe the initial state of H$^-$; rescattering of the
electron from the potential was also investigated, although the
effect was found to be small except for high magnetic field
strengths.  Peters and Delos~\cite{Peter93} gave a semiclassical
analysis of H$^-$ photodetachment in crossed fields and correlated
significant features of the spectrum with closed classical orbits.
Subsequently they derived quantum formulas for this process (using
the zero-range potential model for the initial state) and exhibited
the connection to their predicted  classical closed periodic
orbits~\cite{Peter93b}.  The generalization of Liu et
al.~\cite{Liu96,Liu97,Liu97a} to the case of static electric and
magnetic fields of arbitrary orientation is also based upon the
zero-range potential model.  In all of these works the
electromagnetic field that causes photodetachment is assumed to be
weak and monochromatic.  Also, the photodetachment spectrum is
analyzed numerically only over a very small energy range above
threshold.

In this paper we consider detachment of H$^-$  by a short  laser
pulse in the presence of crossed static electric and magnetic
fields. We present an analytic expression for the final state of the
detached electron taking into account exactly the effects of the
laser field as well as both static fields. The initial state is
described by the solution of the zero-range potential, as in all
other quantum treatments to
date~\cite{Fabr91,Peter93b,Liu96,Liu97,Liu97a}.  We present also an
analytic expression for the photodetachment transition amplitude
that can be used to describe the probabilities of {\it
{multiphoton}} detachment in crossed fields.  In this paper,
however,  our focus is on single photon detachment by short laser
pulses and on the connection between the detached electron wave
packet motion and the predicted  classical closed periodic orbits of
Peters and Delos~\cite{Peter93}.  As noted by Alber and
Zoller~\cite{Alber1991} (in connection with electronic wave packets
in Rydberg atoms), such wave packets ``provide a bridge between
quantum mechanics and the classical concept of the trajectory of a
particle'' and ``the evolution of these wave packets provides
real-time observations of atomic or molecular dynamics.'' We show
this connection for the case of short pulse laser-detached electron
wave packets in crossed static electric and magnetic fields.  In
addition, we show analytically how our short pulse results reduce to
the quantum monochromatic field  results of Fabrikant~\cite{Fabr91}
and Peters and Delos~\cite{Peter93b} in the long pulse limit as well
as the connection between our analytic quantum formulation for the
photodetachment spectrum and those features that we associate with
the predicted classical closed orbits~\cite{Peter93}.  Finally, we
present numerical results in the long pulse limit over a large
energy range above the single photon detachment threshold in order
to demonstrate clearly these manifestations of classical behavior in
our predicted photodetachment spectrum.

This paper is organized as follows:  In Sec. II we present our
theoretical formulation for detachment of H$^-$ by a short laser
pulse in the presence of  crossed static electric and magnetic
fields.  In particular, in this section (with details given in an
Appendix) we present an exact, analytic expression for the wave
function for an electron interacting with both the laser pulse and
the crossed static electric and magnetic fields. We present here
also analytic expressions for the transition probability amplitudes
for both a single laser pulse and a double laser pulse (i.e., two
coherent single pulses separated by a time delay).  In addition, the
long pulse (monochromatic field) limit of our results is presented
and this result is compared with a number of prior works for various
static field cases.  In Sec. III we establish the connection between
the long pulse limit of our results and the closed classical
periodic orbits predicted by Peters and Delos~\cite{Peter93}. In
Sec. IV we present our numerical results, starting first with a
comparison with prior results for the long pulse (monochromatic
field) case and then examining the short pulse case, including the
final state motion of the detached electron wave packets.

%%%%%%%%%%%%%%%%%%%%%%%%%%%%%%%%%%%
%
%
% Theoretical Formulation Section
%
%
%%%%%%%%%%%%%%%%%%%%%%%%%%%%%%%%%%%

\section{\label{sec:level2}Theoretical Formulation}

We consider photodetachment of H$^-$ by one or more short laser
pulses in the presence of crossed static electric and magnetic
fields. In the final state, we assume the detached electron
experiences only the laser and static fields; we ignore final state
interaction of the electron with the residual hydrogen atom. For
weak external fields, this is expected to be a good approximation
for this predominantly single photon process. In this section,  we
first give the $S$-matrix transition amplitude for photodetachment
of H$^{-}$. Then we present an exact quantum mechanical solution to
the time-dependent Schr\"{o}dinger equation for the final state of
the detached electron in both the crossed static electric and
magnetic fields and the time dependent laser pulse. We then use this
result together with S-matrix theory to obtain detachment rates and
cross sections. Atomic units are used throughout this paper unless
otherwise stated.

\subsection{$S$-matrix Transition Amplitude for Photodetachment of H$^{-}$}

We  adopt the Keldysh approximation for the final-state, {\it i.e.},
we neglect the binding potential~\cite{Reiss80}. In this case, the
$S$-matrix transition amplitude from the initial state $\psi_{i}$ to
the final state $\psi_{f}$ is  given by
\begin{equation}
 S_{fi} = - i\int_{-\infty
}^{\infty}dt^{\prime}\left\langle \psi_{f}(\mathbf{p},t^{\prime})|
V_I(t^{\prime})|\psi_{i}(\mathbf{p},t^{\prime}) \right\rangle,
\label{Stime0}
\end{equation}
where $V_I$ represents the laser-electron interaction and the
bracket $\left\langle {}\right\rangle $ stands for integration over
momentum space. For the zero range potential for which the bound
state wave function has the form $e^{-\kappa r}/r$, the $S$-matrix
element in Eq.~(\ref{Stime0}) can be shown to be
gauge-invariant~\cite{Gao90,Niki66}. Such a bound state wave
function can be used to represent the weakly bound electron of
H$^{-}$. We use that of Ohmura and Ohmura~\cite{Ohmu60}, which in
momentum space is given by
\begin{equation}
\psi _{i}(\mathbf{p},t)=\frac{C_{i}}{\sqrt{2\pi
}}\frac{e^{-i\varepsilon _{i}t}}{%
p^{2}/2-\varepsilon _{i}},  \label{initialwave}
\end{equation}%
where $C_{i}$ is a normalization constant and $\varepsilon _{i}$ is
the initial state energy.  Using the variational results of
Ref.~\cite{Ohmu60} and effective range theory for a weakly bound
$s$-electron~\cite{Beth50}, one obtains~\cite{Du88} $C_{i}=0.31552$
and $\varepsilon _{i}=-0.027751$ a.u..  The gauge-invariant
$S$-matrix transition amplitude for H$^-$ detachment is then given
by (cf. Eq.~(27) of Ref.~\cite{Gao90})
\begin{equation}
\left( S_{fi}\right) _{k_{x}k_{y}n_{z}}=i\int_{-\infty
}^{\infty}dt^{\prime}\left\langle \psi _{f}(\mathbf{p},t^{\prime
})|C_{i}/\sqrt{2\pi}\right\rangle e^{-i\varepsilon _{i}t^{\prime}}.
\label{Stime}
\end{equation}

\subsection{The Final State  Wave Function}
\begin{figure}
\centering
\includegraphics[width=12cm]{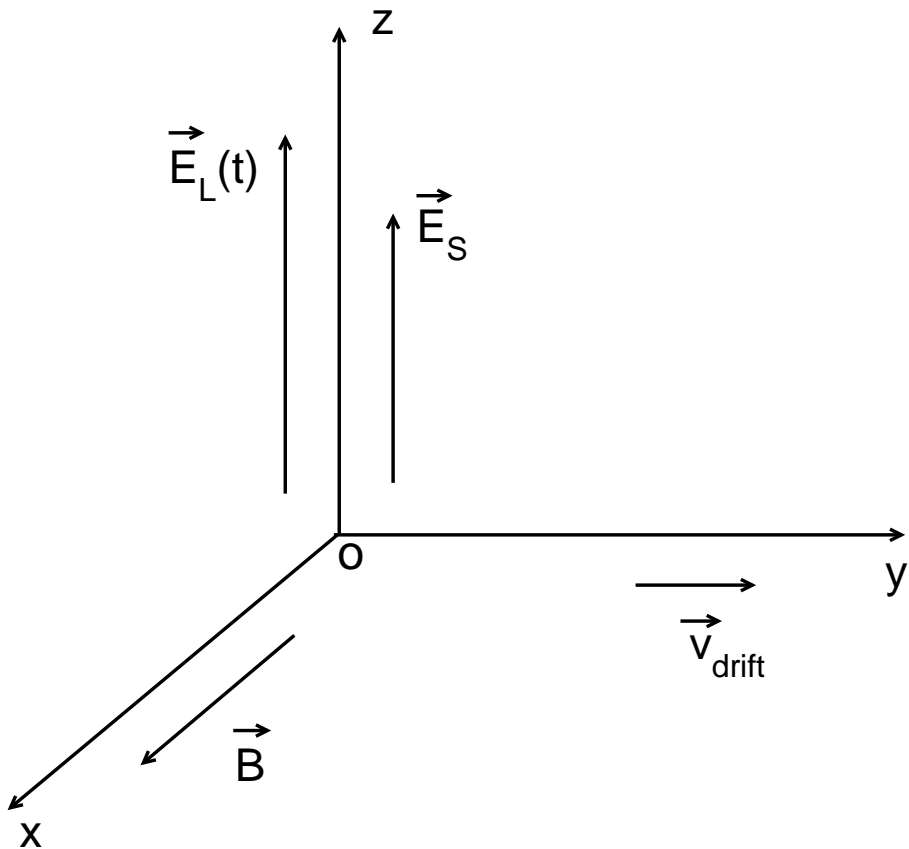}\\
    \caption{Geometrical arrangement of fields in photodetachment of $H^-$ by a linearly polarized laser (with electric field $\bf{E_L}$) in the presence of crossed static electric ($\bf{E_S}$) and magnetic ($\bf{B}$) fields.  Both the laser and the static electric fields point along the $z$-axis.  As indicated, the drift motion of the detached electron is along the $y$-axis.}
    \label{figure1}
\end{figure}
 In order to calculate the $S$-matrix transition amplitude in
 Eq.~(\ref{Stime}), we  present in this section an analytical expression
 for the final state wave function $\psi _{f}$. As aforementioned,
 we neglect the binding potential after detachment. Therefore,
 $\psi _{f}$ is actually a Volkov-type wave function that describes a
 free electron  moving  in the combined field of the crossed static
 electric and magnetic fields and  the time-dependent electric field
 associated with the short laser pulse. In Fig.~\ref{figure1} we illustrate
 the configuration of the
external fields in which the detached electron moves:  the uniform
static magnetic field defines the $x$ axis and the static electric
field defines the $z$ axis, i.e.,
\begin{eqnarray}
&&\mathbf{B}=B\hat{\mathbf i}\\
&&\mathbf{E_{S}}=E_{S}\hat{\mathbf k}. \label{staticfields}
\end{eqnarray}%
We assume that each laser pulse has the following general form,
\begin{equation}
\mathbf{E_L}(t)=E_{0}e^{-\alpha ^{2}\left( t-\tau \right)
^{2}}\sin \left( \omega t+\beta \right) \hat{\mathbf k},
\label{laserpulse}
\end{equation}%
where $\omega $ is the laser frequency, $\tau $ is the time delay
with respect to $t=0$, and $\beta $ is a (generally constant)
phase. The duration of the laser pulse is defined to be the full
width at half maximum (FWHM) of the laser intensity, and is given
by
\begin{equation}
T_{p}=\sqrt{2\ln 2}/\alpha.  \label{Tpulse}
\end{equation}%
We introduce the vector potentials for the magnetic field and the
laser field respectively, as follows:
\begin{eqnarray}
&&\mathbf{A_B}= -zB\hat{\mathbf j}, \\
&&\mathbf{A_L}(t)= -c\int_{-\infty}^{t}\mathbf{E_L}(t^{\prime
})dt^{\prime},  \label{vectpot}
\end{eqnarray}
where $c$ is the speed of light in vacuum.

The final state wave function for the detached electron is obtained
as the solution of the time-dependent Schr\"{o}dinger equation
(TDSE) in momentum space,
\begin{equation}
i\frac{\partial}{\partial t}\psi _{f}^{(p)}(\mathbf{p},t)=H\psi
_{f}^{(p)}(\mathbf{p},t), \label{tdsemom}
\end{equation}
in which the Hamiltonian $H$ is given by
\begin{eqnarray}
H(\mathbf{p},t)  &= & \frac{1}{2}\left[ \mathbf{p}+\frac{1}{c}%
\left( \mathbf{A}_{L}+\mathbf{A}_{B}\right) \right]^{2}+
\mathbf{\hat{r}} \cdot \mathbf{E}_S \\ \label{hamiltonianmom0}
&=&-\frac{1}{2}\omega _{c}^{2}\frac{\partial
^{2}}{\partial p_{z}^{2}}%
-i\omega _{c}\left( p_{y}-\frac{E_{S}}{\omega _{c}}\right) \frac{\partial}{%
\partial p_{z}}+\frac{1}{2}p_{z}^{2}+\frac{1}{c}p_{z}A_{L}(t)  \nonumber \\
&&+\frac{1}{2}p_{x}^{2}+\frac{1}{2}p_{y}^{2}+\frac{1}{2c^{2}}A_{L}^{2}(t),
\label{hamiltonianmom}
\end{eqnarray}
where  $\omega _{c}=B/c$ is the cyclotron frequency.

It can be shown that Eq.~(\ref{tdsemom}) has  an exact analytical
solution. The details of the derivation are presented in
Appendix~\ref{appendixA}.  The final expression of the solution is
given by
\begin{eqnarray}
\psi _{f}^{(p)}(\mathbf{p},t)&=&\delta (p_{x}-k_{x})\delta
(p_{y}-k_{y})\exp \left[ -i\varepsilon _{f}t-if(t)\right] \nonumber \\
&\times& \omega _{c}^{-1/4} g_{n_{z}}\left(\sqrt{2}\zeta
_{p_{z}}\right) \exp \left[ -ib(k_{y},t)\sqrt{2}\zeta
_{p_{z}}\right],  \label{finalwvmom}
\end{eqnarray}
in which $g_{n_{z}}\left( x\right)$  is defined by %%
\begin{equation}
g_{n_{z}}\left( x \right)= \frac{1}{\sqrt{2^{n_{z}}n_{z}!\sqrt{\pi
}}}e^{-x^{2}/2}H_{n_{z}}(x), \label{gnzetay}
\end{equation}
where $H_{n_{z}}(x)$ is the  $n_z\it{th}$ Hermite polynomial. In
Eq.~(\ref{finalwvmom}) we have also defined
\begin{eqnarray}
 && \varepsilon _{f} = \frac{1}{2}\left( k_{x}^{2}+k_{y}^{2}\right)
+\varepsilon _{n_{z}}-\frac{1}{2}\omega _{c}\zeta _{k_{y}}^{2},  \nonumber \\
 && \hspace{0.48cm}=\frac{1}{2}k_{x}^{2}+\frac{E_{S}}{\sqrt{\omega _{c}}}\zeta
_{k_{y}}+\varepsilon _{n_{z}}+\frac{1}{2}\frac{%
E_{S}^{2}}{\omega _{c}^{2}},  \label{epsf} \\
&& f(t)= \frac{1}{\sqrt{\omega _{c}}}\zeta _{k_{y}}\xi (t)+\frac{1}{2c^{2}}%
\int_{-\infty}^{t}A_{L}^{2}(t^{\prime})dt^{\prime}-\int_{-\infty
}^{t}L(t^{\prime})dt^{\prime},  \label{ft} \\
&& b(k_{y},t)= \zeta _{k_{y}}-\dot{\xi}(t)/\sqrt{\omega _{c}^{3}},
\label{bt}
\end{eqnarray}
where  the arguments $\zeta _{p_{z}}$ and $\zeta _{k_{y}}$ are given
by
\begin{eqnarray}
&&\zeta _{p_{z}}(t)= \left[ p_{z}-\xi (t)\right] /\sqrt{2\omega
_{c}},
\label{zetapz}  \\
&&\zeta _{k_{y}}= \frac{1}{\sqrt{\omega _{c}}}\left( k_{y}-\frac{E_{S}}{%
\omega _{c}}\right),  \label{zetaky1}
\end{eqnarray}
and the energy of the $n_zth$ Landau
level $\varepsilon _{n_{z}}$  in Eq.~(\ref{epsf})  is given by
\begin{equation}
\varepsilon _{n_{z}} = \left( n_{z}+\frac{1}{2}\right) \omega _{c}.
\label{epsnz}
\end{equation}

In Eq.~(\ref{ft}), $L(t)$  and $\xi(t)$  are functions
 related to the vector potential of the short laser pulse.
  $L(t)$  is  defined  as
\begin{equation}
L(t) =  \frac{1}{2\omega _{c}^{2}}\dot{\xi}^{2}(t)-\frac{1}{2}\xi ^{2}(t)-%
\frac{1}{c}A_{L}(t)\xi (t),  \label{tdsepsiz3b}
\end{equation}
while $\xi (t)$ satisfies the
following differential equation:
\begin{equation}
\ddot{\xi}(t)+\omega _{c}^{2}\xi (t)=-\frac{\omega
_{c}^{2}}{c}A_{L}(t), \label{odeksi}
\end{equation}
where $\ddot{\xi}(t)$ denotes the second derivative of $\xi (t)$.
We present the exact solution for $\xi(t)$ in Appendix
\ref{appendixB}. However, in the long pulse case ($\alpha/\omega \ll 1$),
 a simplified  expression for $\xi(t)$ can be obtained as
\begin{equation}
\xi (t) \simeq a\left( \omega \right) e^{-\alpha ^{2}(t-\tau
)^{2}}\cos \left( \omega t+\beta \right),  \label{longksi}
\end{equation}
where we have defined
\begin{equation}
a\left( \omega \right) =\frac{E_{0}\,\omega _{c}^{2}}{\omega \left(
\omega ^{2}-\omega _{c}^{2}\right)}.  \label{constA}
\end{equation}
In this case, the first derivative, $\dot{\xi}(t)$,  is thus given by
\begin{equation}
\dot{\xi}(t)\simeq -\omega a\left( \omega \right) e^{-\alpha
^{2}(t-\tau )^{2}}\sin \left( \omega t+\beta \right).
\label{longksidot}
\end{equation}

 In order to investigate  wave  packet dynamics, it is useful to
derive an expression for the final state wave function in which its
$z$ component is given in coordinate space. This is achieved by
taking the Fourier transform of Eq.~(\ref{finalwvmom}) with respect
to $p_{z}$, i.e.,
\begin{equation}
\psi _{f}^{(z)}(p_{x},p_{y},z,t) \equiv \psi
_{k_{x},k_{y},n_z}^{(z)}(p_{x},p_{y},z,t) =\frac{1}{\sqrt{2\pi
}}\int_{-\infty}^{\infty}dp_{z}\psi
_{f}^{(p)}(\mathbf{p},t)e^{izp_{z}}. \label{finalwvrealz}
\end{equation}
Changing the integration variable to $\zeta _{p_{z}}$ (cf.
Eq.~(\ref{zetapz})), we obtain
\begin{eqnarray}
\psi _{f}^{(z)}(p_{x},p_{y},z,t) &=&\delta (p_{x}-k_{x})\delta
(p_{y}-k_{y})i^{n_{z}}\omega _{c}^{1/4}g_{n_{z}}\left[ \sqrt{\omega _{c}}%
z-b(k_{y},t)\right]  \nonumber \\
&&\times \exp \left[ iz\xi (t)-i\varepsilon _{f}t-if(t)\right],
\label{finalwvrealz1}
\end{eqnarray}
where we have made use of Eqs.~7.388(2) and~7.388(4) in
Ref.~\cite{Gradsh}.

\subsection{$S$-matrix Amplitude for Photodetachment of H$^{-}$}

  %%%%%%%%%%%%%%%%%%%%%%%%%%%%%%%%%%%%%%%%%%%%%%%
  %%%%%%%%%%%%%%%%%%%%%%%%%%%%%%%%%%%%%%%%%%%%%%%
   %%%%%%%%%%%%%%%%%%%%%%%%%%%%%%%%%%%%%%%%%%%%%%%

In order to examine the motion of the detached electron wave packet
in crossed $\mathbf{E}$ and $\mathbf{B}$ fields, we define in
analogy to Eq.~(23) of Ref.~\cite{Wang95} a time-dependent
transition amplitude $R_{fi}(t)$ from the initial state to the final
state $(k_{x},k_{y},n_{z})$:
\begin{eqnarray}
\left( R_{fi}(t)\right)_{k_{x}k_{y}n_{z}} &=&i\frac{C_{i}}{\sqrt{2\pi}}%
\sqrt{2\omega _{c}}\int_{-\infty}^{t}dt^{\prime}e^{i\varepsilon
_{fi}t^{\prime}+if(t^{\prime})}  \nonumber \\
&&\times \int_{-\infty}^{\infty} \omega _{c}^{-1/4}
g_{n_{z}}\left(\sqrt{2}\zeta _{p_{z}}\right) \exp \left[ i
b(k_{y},t^{\prime}) \sqrt{2}\zeta _{p_{z}}\right] d\zeta _{p_{z}},
\end{eqnarray}
where $\varepsilon _{fi}= \varepsilon _{f}-\varepsilon _{i}$,
$b(k_{y},t^{\prime})$ is given by Eq.~(\ref{bt}), and where we have
used Eq.~(\ref{finalwvmom}) for the final state wave function in
Eq.~(\ref{Stime}). Using  Eqs.~7.388(2) and~7.388(4) in
Ref.~\cite{Gradsh} to carry out the integration over $\zeta
_{p_{z}}$, we obtain %%
\begin{equation}
   \left( R_{fi}(t)\right) _{k_{x}k_{y}n_{z}} =
   i^{n_{z}+1}C_{i}\int_{-\infty}^{t}dt^{\prime}e^{i\varepsilon
_{fi}t^{\prime}+if(t^{\prime})}\omega _{c}^{1/4} g_{n_{z}}\left[
b(k_{y},t^{\prime})\right]. \label{momSmatr}
\end{equation}
Note that in the limit of  $t\rightarrow \infty$,  $R_{fi}(t)$ reduces to the
 $S$-matrix transition amplitude (\ref{Stime}), {\it i.e.}:
\begin{equation}
\left( S_{fi}\right) _{k_{x}k_{y}n_{z}} = \lim _{t\rightarrow\infty}
\left( R_{fi}(t)\right) _{k_{x}k_{y}n_{z}}.
\end{equation}

In principle, with this analytical  $S$-matrix amplitude one can
readily calculate the total and multiphoton transition rates, as
done for H$^-$ detachment in a static electric field in
Ref.~\cite{Gao90}. However, in the present paper, we restrict our
consideration to the one-photon detachment process (Note that there are
still many cycles in the laser pulses that we will consider in this work).  Our
analytical results  facilitate easy comparison with some other
previous results. Consequently, we evaluate Eq.~(\ref{momSmatr})
only to first order in the laser electric field strength $E_{0}$,
i.e., we employ the following approximations:
\begin{eqnarray*}
e^{if(t^{\prime})} &\simeq &1+if(t^{\prime}) \\
&\simeq &1+i\frac{1}{\sqrt{\omega _{c}}}\zeta _{k_{y}}\xi
(t^{\prime}),
\end{eqnarray*}
\[
g_{n_{z}}\left[ b(k_{y},t^{\prime}\right] \simeq g_{n_{z}}
(\zeta _{k_{y}})+g_{n_{z}}^{\prime}(\zeta _{k_{y}})%
\left[ -\frac{\dot{\xi}(t^{\prime})}{\sqrt{\omega _{c}^{3}}}\right]
,
\]
where $g_{n_{z}}^{\prime}(\zeta _{k_{y}})$ stands for the derivative
of $g_{n_{z}}(\zeta _{k_{y}})$. Thus, to  first order in $E_{0}$,
the time-dependent transition amplitude is given by
\begin{eqnarray}
\left( R_{fi}^{(1)}(t)\right) _{k_{x}k_{y}n_{z}}
&=&i^{n_{z}+1}C_{i}\omega _{c}^{1/4}g_{n_{z}} (\zeta
_{k_{y}})\int_{-\infty}^{t}dt^{\prime}e^{i\varepsilon
_{fi}t^{\prime}}  \nonumber \\
&&-i^{n_{z}+1}C_{i}\omega _{c}^{1/4}g_{n_{z}}^{\prime}(\zeta
_{k_{y}})\int_{-\infty}^{t}dt^{\prime}e^{i\varepsilon _{fi}t^{\prime
}}\frac{\dot{\xi}(t^{\prime})}{%
\sqrt{\omega _{c}^{3}}}  \nonumber \\
&&-i^{n_{z}}C_{i}\zeta _{k_{y}}\omega _{c}^{1/4}g_{n_{z}}(\zeta
_{k_{y}})\int_{-\infty}^{t}dt^{\prime}e^{i\varepsilon
_{fi}t^{\prime}}\frac{%
\xi (t^{\prime})}{\sqrt{\omega _{c}}}.  \label{Smatrixfinal2}
\end{eqnarray}
Note that, as usual, the first term in Eq.~(\ref{Smatrixfinal2})
   does not contribute to the photodetachment process
(since for $t\rightarrow \infty$, the only contributions are for
$\varepsilon_{fi}\rightarrow 0$); hence  this term is discarded in
the following discussion.

In the long pulse approximation, with the help of Eqs.~(\ref%
{longksi}) and~(\ref{longksidot}), one can show that
\begin{eqnarray}
\left( R_{fi}^{(1)}(t)\right) _{k_{x}k_{y}n_{z}}
&=&i^{n_{z}+1}C_{i}
g_{n_{z}}^{\prime}(\zeta _{k_{y}})\frac{\omega a(\omega)}{\omega _{c}^{5/4}}%
\int_{-\infty}^{t}dt^{\prime}e^{-\alpha ^{2}(t^{\prime}-\tau
)^{2}+i\varepsilon _{fi}t^{\prime}}\sin \left( \omega t^{\prime
}+\beta
\right)   \nonumber \\
&&-i^{n_{z}}C_{i}\zeta _{y}g_{n_{z}}(\zeta _{k_{y}})\frac{(\omega)}
{\omega _{c}^{1/4}}\int_{-\infty}^{t}dt^{\prime}e^{-\alpha
^{2}(t^{\prime}-\tau )^{2}+i\varepsilon _{fi}t^{\prime}}\cos \left(
\omega t^{\prime}+\beta \right),
\end{eqnarray}
which reduces to
\begin{eqnarray}
\left( R_{fi}^{(1)}(t)\right) _{k_{x}k_{y}n_{z}}
&=&-i^{n_{z}}C_{i}\frac{a(\omega)}{2%
\omega _{c}^{5/4}}\left[ \omega g _{n_{z}}^{\prime}(\zeta
_{k_{y}})+\omega _{c}\zeta _{y}g _{n_{z}}(\zeta _{k_{y}})\right]
\nonumber \\
&&\times \int_{-\infty}^{t}dt^{\prime}e^{-\alpha ^{2}(t^{\prime
}-\tau )^{2}+i\varepsilon _{fi}t^{\prime}-i\omega t^{\prime}-i\beta
} \label{fstSm}
\end{eqnarray}
if we neglect the emission process (i.e., if we discard terms
involving $e^{+i\omega t^{\prime}}$).

The integration over $t^{\prime}$ in Eq.~(\ref{fstSm}) can be
carried out analytically:
    %% %%
     %%
\begin{eqnarray*}
&&\int_{-\infty}^{t}dt^{\prime}e^{-\alpha ^{2}(t^{\prime}-\tau
)^{2}+i\varepsilon _{fi}t^{\prime}-i\omega t^{\prime}-i\beta} \\
&=&\frac{\sqrt{\pi}}{2\alpha}\left\{%TCIMACRO{\func{erf}}%
%BeginExpansion
\mathop{\rm erf}%
%EndExpansion
\left[ \alpha \left( t-\tau \right) -i\frac{\varepsilon
_{fi}-\omega
}{2\alpha}%
\right] +1\right\}  \\
&&\times \exp \left[ -\frac{\left( \varepsilon _{fi}-\omega \right) ^{2}}{%
4\alpha ^{2}}+i\left( \varepsilon _{fi}-\omega \right) \tau
-i\beta \right].
\end{eqnarray*}
Thus, for the single laser pulse in  Eq.~(\ref{laserpulse}), the
first-order time-dependent transition amplitude in the long pulse
approximation is given by%
\begin{eqnarray}
\left( R_{fi}^{(1)}(t)\right) _{k_{x}k_{y}n_{z}}^{\text{sgl}}
&=&-i^{n_{z}}C_{i}\frac{\pi a(\omega)%
}{2\omega _{c}^{5/4}}\left[ \omega g_{n_{z}}^{\prime}(\zeta
_{k_{y}})+\omega _{c}\zeta _{y}
g_{n_{z}}(\zeta _{k_{y}})\right]   \nonumber \\
&&\times D_{\text{sgl}}(\varepsilon _{fi},t)
\delta_{\alpha}(\varepsilon _{fi}-\omega),
\label{Smatrixfinaltime1}
\end{eqnarray}
where we have defined
\begin{equation}
D_{\text{sgl}}(\varepsilon _{fi},t)=e^{i\left( \varepsilon
_{fi}-\omega \right)
\tau -i\beta}\left[ 1+%
%TCIMACRO{\func{erf}}%
%BeginExpansion
\mathop{\rm erf}%
%EndExpansion
\left[ \alpha \left( t-\tau \right) -i\frac{\varepsilon
_{fi}-\omega
}{2\alpha}%
\right] \right],  \label{Dsglt}
\end{equation}
and have also introduced  the
quasi-$\delta$-function~\cite{Wang95}, %%
\begin{equation}
\delta_{\alpha}(\varepsilon _{fi}-\omega) = \left( 2 \pi^{1/2}
\alpha\right)^{-1} \exp \left[ -\frac{\left( \varepsilon
_{fi}-\omega \right) ^{2}}{4\alpha ^{2}}\right].
\end{equation}
In the limit that our finite laser pulse becomes a monochromatic
plane wave, the quasi-$\delta$-function becomes the usual Dirac
$\delta$-function, %%
\begin{equation}
\delta(\varepsilon _{fi}-\omega) = \lim_{\alpha\rightarrow
0}\delta_{\alpha}(\varepsilon _{fi}-\omega). \label{quasidelta1}
\end{equation}

Taking the limit $t\rightarrow +\infty$, we obtain from
Eq.~(\ref{Smatrixfinaltime1}) the following analytical expression
for the $S$-matrix amplitude for the case of a single, finite
laser pulse: %%
\begin{eqnarray}
\left( S_{fi}^{(1)}\right) _{k_{x}k_{y}n_{z}}^{\text{sgl}}
&=&-i^{n_{z}}C_{i}\frac{\pi a(\omega)}{\omega _{c}^{5/4}}\left[
\omega g _{n_{z}}^{\prime}(\zeta _{k_{y}})+\omega _{c}\zeta _{y}g
_{n_{z}}(\zeta _{k_{y}})\right]   \nonumber \\
&&\times e^{i\left( \varepsilon _{fi}-\omega \right) \tau -i\beta
}\delta_{\alpha}(\varepsilon _{fi}-\omega), \label{Smatrixfinalinf}
\end{eqnarray}
where we have used the fact that $%
%TCIMACRO{\func{erf}}%
%BeginExpansion
\mathop{\rm erf}%
%EndExpansion
(\infty +iy)=1$ for any finite real number $y$.

\subsection{Detached Electron Wave  Packet}
We may obtain the detached electron wave packet probability
amplitude as a sum over all final states of the product of the
time-dependent transition amplitude   for transition  to the final
state $(k_x, k_y,n_z)$ at a particular time $t$
[Eq.~(\ref{Smatrixfinaltime1})] and the wave function
[Eq.~(\ref{finalwvrealz1})] for that state (cf. Sec. II. D of
Ref.~\cite{Wang95}):
\begin{equation}
\psi_{\text{WP}}(p_{x},p_{y},z,t)=\sum_{n_{z}=0}^{\infty
}\int_{-\infty}^{\infty}dk_{x}\int_{-\infty}^{\infty}dk_{y}\psi
_{k_{x},k_{y},n_{z}}^{(z)}(p_{x},p_{y},z,t)\left(
R_{fi}^{(1)}(t)\right) _{k_{x}k_{y}n_{z}}.
\end{equation}%
By using Eqs.~(\ref{finalwvrealz1}) and~(\ref{Smatrixfinaltime1}),
the wave  packet for the single laser pulse~(\ref{laserpulse})  is
given by
\begin{eqnarray}
\psi^{\text{sgl}}_{\text{WP}}(p_{x},p_{y},z,t)
&=&-C_{i}\frac{\pi a(\omega)}{2 \omega _{c}}%
\sum_{n_{z}=0}^{\infty}\left( -1\right) ^{n_{z}}\exp \left[ iz\xi
-i\varepsilon _{f}^{\prime}t-i\frac{1}{\sqrt{\omega _{c}}}\zeta _{p_{y}}\xi %
\right]   \nonumber \\
&&\times g_{n_{z}}\left[ \sqrt{\omega _{c}}z-b(p_{y},t)%
\right] \left[ \omega g_{n_{z}}^{\prime}(\zeta _{p_{y}})+\omega
_{c}\zeta _{y}g_{n_{z}}(\zeta _{p_{y}})\right]   \nonumber \\
&&\times D_{\text{sgl}}(\varepsilon _{fi}^{\prime
},t)\delta_{\alpha}(\varepsilon _{fi}^{\prime}-\omega)
\label{wavepacket1}
\end{eqnarray}%
where $\zeta _{p_{y}}$ and $b(p_{y},t)$ are defined by
Eqs.~(\ref{zetaky1}) and~(\ref{bt}) respectively, $\varepsilon
_{fi}^{\prime} =\varepsilon_f^{\prime} - \varepsilon_i$, and
$\varepsilon_f^{\prime}$ is given by Eq.~(\ref{epsf}) with $\zeta
_{k_{y}}$ replaced by $\zeta _{p_{y}}$.

\subsection{$S$-matrix and Wave Packet Amplitudes for the Double Pulse Case}
We consider here the case that there are two laser pulses of the
form of Eq.~(\ref{laserpulse}), with the second one delayed with
respect to the first   by a time interval $\tau$ and having a
relative phase of $\beta$, i.e.,
\begin{equation}
\mathbf{E}_{\mathbf{L}}^{\mathbf{{dbl}}}(t)=  E_{0}\left[
e^{-\alpha ^{2}t^{2}}\sin \left( \omega t\right) +  e^{-\alpha
^{2}\left( t-\tau \right) ^{2}}\sin \left( \omega t+\beta \right)
\right] \hat{\mathbf k}.
   \label{laserpulsedbl}
\end{equation}
To  first order in $E_0$, it is easy to show that for the double
laser pulse case,  the time-dependent
transition amplitude is given by%
\begin{eqnarray}
\left( R_{fi}^{(1)}(t)\right) _{k_{x}k_{y}n_{z}}^{\text{dbl}}
&=&-i^{n_{z}}C_{i}\frac{\pi a(\omega)}{2 \omega _{c}^{5/4}}\left[
\omega g _{n_{z}}^{\prime}(\zeta _{k_{y}})+\omega _{c}\zeta _{y}g
_{n_{z}} (\zeta _{k_{y}})\right]   \nonumber \\
&&\times D_{\text{dbl}}(\varepsilon _{fi},t)
\delta_{\alpha}\left(\varepsilon _{fi}-\omega \right),
   \label{crsdbl}
\end{eqnarray}
where the function $D_{\text{dbl}}(\varepsilon _{fi},t)$ is given
by
\begin{equation}
D_{\text{dbl}}(\varepsilon _{fi},t)=1+%
%TCIMACRO{\func{erf}}%
%BeginExpansion
\mathop{\rm erf}%
%EndExpansion
\left[ \alpha t-i\frac{\varepsilon _{fi}-\omega}{2\alpha}\right]
+e^{i\left(
\varepsilon _{fi}-\omega \right) \tau -i\beta}\left\{1+%
%TCIMACRO{\func{erf}}%
%BeginExpansion
\mathop{\rm erf}%
%EndExpansion
\left[ \alpha \left( t-\tau \right) -i\frac{\varepsilon
_{fi}-\omega
}{2\alpha}%
\right] \right\}.  \label{Ddblt}
\end{equation}%
When $t\rightarrow \infty $, the above formula reduces to
\begin{eqnarray}
\left( S_{fi}^{(1)}\right) _{k_{x}k_{y}n_{z}}^{\text{dbl}}
&=&\lim_{t\rightarrow \infty}\left( R_{fi}^{(1)}(t)\right)
_{k_{x}k_{y}n_{z}}^{\text{dbl}}  \nonumber \\
&=&-i^{n_{z}}C_{i}\frac{\pi a(\omega)}{\omega _{c}^{5/4}}\left[
\omega g _{n_{z}}^{\prime}(\zeta _{k_{y}})+\omega _{c}\zeta _{y}g
_{n_{z}} (\zeta _{k_{y}})\right]   \nonumber \\
&&\times \left[ 1+e^{i\left( \varepsilon _{fi}-\omega \right) \tau -i\beta}%
\right] \delta_{\alpha}\left(\varepsilon _{fi}-\omega \right).
\label{crsdbl0}
\end{eqnarray}
   The wave  packet amplitude for
the double laser pulse case is correspondingly given by
\begin{eqnarray}
\psi _{\text{WP}}^{\text{dbl}}(p_{x},p_{y},z,t) &=&-C_{i}\frac{\pi a(\omega) %
}{2 \omega _{c}}\sum_{n_{z}=0}^{\infty}\left( -1\right) ^{n_{z}}\exp
\left[ iz\xi -i\varepsilon _{f}^{\prime}t-i\frac{1}{\sqrt{\omega
_{c}}}\zeta _{p_{y}}\xi \right]   \nonumber \\
&&\times g _{n_{z}} \left[ \sqrt{\omega _{c}}z-b(p_{y},t)%
\right] \left[ \omega g _{n_{z}}^{\prime}(\zeta _{p_{y}})+\omega
_{c}\zeta _{y}g _{n_{z}} (\zeta _{p_{y}})\right]   \nonumber \\
&&\times D_{\text{dbl}}(\varepsilon _{fi}^{\prime
},t)\delta_{\alpha}\left(\varepsilon _{fi}^{\prime}-\omega
\right). \label{wvpkdble}
\end{eqnarray}

\subsection{Photodetachment Cross Section}

The  transition probability to a particular final state $\left(
k_{x},k_{y},n_{z}\right) $ is given by
\begin{equation}
P_{k_{x}k_{y}n_{z}}=\left| \left(
S_{fi}\right) _{k_{x}k_{y}n_{z}}\right| ^{2}, \label{wrate}
\end{equation}
and the total photodetachment probability is calculated by
integrating over all  final states,
\begin{equation}
P=\sum\limits_{n_{z}=0}^{\infty}\int_{-\infty}^{\infty
}dk_{x}\int_{-\infty}^{\infty}dk_{y}P_{k_{x}k_{y}n_{z}}.
\label{transrate}
\end{equation}
For an infinitely long,  monochromatic beam, the probability $P$ is
proportional to time, $t$. In this case, it does not make sense to
talk about the total transition probability. Instead, one normally
considers the total transition rate,  $W$, which is given
by~\cite{Reiss80}
 \begin{equation}
   W = \lim_{t\rightarrow\infty} \frac{1}{t} P.
 \end{equation}
The total photodetachment cross section is obtained by dividing
the total  photodetachment rate $W$  by the photon flux $F$ (the
number of photons  per unit area per unit time):
\begin{equation}
\sigma_{pw} =\frac{W}{F}, \label{crsdef0}
\end{equation}%
where `{\it pw}' stands for the monochromatic  plane wave case.

For the short laser pulse case, it does not make sense to
talk about a transition rate since the transition probability is not
simply proportional to time $t$. In addition, the photon flux $F$ is
not well defined.  Nevertheless, it is possible to renormalize the total probability for detachment by a short laser pulse in such a way that the renormalized probability reduces, in the limit of an infinitely long pulse, to the usual formula for the photodetachment cross section.  Since the renormalized probability will have the dimensions of area, we denote it as an {\it effective photodetachment cross section}, $\sigma$. To derive this effective cross section,  one uses the time duration of the laser pulse as the unit of time. One calculates the total  photodetachment    probability $P$ during the laser pulse duration and the
total number of photons per unit area (i.e., the photon density),  $\Sigma$, during the laser pulse duration.   Then an effective photodetachment cross section,   $\sigma$,   may be defined as
  \begin{equation}
   \sigma =\frac{P}{\Sigma}. \label{crsdef}
  \end{equation}
Clearly $\sigma$ defined in this way  has the
dimensions of a cross section.  In the rest of this paper, $\sigma$
should be understood to be this effective photodetachment cross section, i.e.,
calculated according to Eq.~(\ref{crsdef}). We shall show below that  this $\sigma$
for the short laser pulse case reduces in the limit
$\alpha\rightarrow0$ (cf. Eq.~(\ref{laserpulse}))
to the usual photodetachment cross section for a monochromatic plane wave.

    It has been shown by Wang and Starace~\cite{Wang95} that for a
single Gaussian pulse defined by Eq.~(\ref{laserpulse}) with $\tau
=\beta=0$,   the photon density $\Sigma$ is given by the following
formula:
\begin{equation}
\Sigma_{\text{sgl}}=\frac{cE_{0}^{2}}{8\pi \omega} \frac{\sqrt{2\pi}}{2\alpha}.
  \label{flux0}
\end{equation}
For the double pulse case (cf.  Eq.~(\ref{laserpulsedbl})), $\Sigma$
is correspondingly given by
\begin{equation}
\Sigma_{\text{dbl}}=\frac{cE_{0}^{2}}{8\pi \omega}\frac{\sqrt{2\pi}}{\alpha}\left[ 1+\cos \beta \exp (-\alpha ^{2}\tau
^{2}/2)\right]. \label{flux2}
\end{equation}
   Taking $\beta $ and $\tau $ to be zero in Eq.~(\ref%
{Smatrixfinalinf}), and using Eqs.~(\ref{constA}) and
(\ref{wrate})-(\ref{flux0}), we have for the photodetachment cross
section of H$^-$ by  a single pulse of the form of
Eq.~(\ref{laserpulse}):
\begin{eqnarray}
\sigma ^{(1)} &=&\frac{4\pi ^{2}C_{i}^{2}\omega _{c}^{2}}{c\omega
(\omega ^{2}-\omega _{c}^{2})^{2}}\sum\limits_{n_{z}=0}^{\infty
}\int_{-\infty}^{\infty}d\zeta _{k_{y}}\left[ \zeta _{k_y}\omega
_{c}g_{n_{z}}\left( \zeta _{k_{y}}\right) +\omega g_{n_{z}}^{\prime
}\left( \zeta _{k_{y}}\right) \right] ^{2}
\nonumber \\
&&\times \int_{-\infty}^{\infty}dk_{x}
\overline{\delta}_{\alpha}\left(\varepsilon _{fi}-\omega \right),
\label{crs1st}
\end{eqnarray}
where we have employed a second quasi-$\delta$
function~\cite{Wang95},
   \begin{equation}
    \overline{\delta}_{\alpha}\left(\varepsilon _{fi}-\omega
\right) = \frac{1}{\alpha \sqrt{2\pi}}%
\exp \left[ -\frac{\left( \varepsilon _{fi}-\omega \right) ^{2}}{2\alpha ^{2}}%
\right], \label{quasidelta2}
   \end{equation}
   which reduces to the usual Dirac $\delta$-function  in the limit
   of a monochromatic plane wave, i.e.,
   \begin{equation}
    \delta\left(\varepsilon _{fi}-\omega
\right) = \lim_{\alpha\rightarrow 0}
\overline{\delta}_{\alpha}\left(\varepsilon _{fi}-\omega \right).
   \end{equation}
   Note that
\begin{equation}
\varepsilon _{fi}-\omega =\frac{1}{2}\left[ k_{x}^{2}+Q\left(
\zeta _{k_{y}}\right) \right], \label{finaleps}
\end{equation}
in which we have defined (cf. Eq.~\ref{epsf})
\begin{equation}
Q\left( \zeta _{k_{y}}\right) =\frac{2E_{S}}{\sqrt{\omega
_{c}}}\left( \zeta _{k_{y}}+\zeta _{\min}\right), \label{Kzetaky}
\end{equation}
   where
\begin{equation}
\zeta _{\min}=\frac{\sqrt{\omega _{c}}}{E_{S}}\left[ (n_{z}+\frac{1}{2}%
)\omega _{c}+\frac{E_{S}^{2}}{2\omega _{c}^{2}}-\varepsilon
_{i}-\omega \right].  \label{zetamin}
\end{equation}
The integration over $k_{x}$ in Eq.~(\ref{crs1st}) has an
analytical
result when $%
Q\left( \zeta _{k_{y}}\right) \geq 0$. The result is
\begin{eqnarray}
&&\int_{-\infty}^{\infty}dk_{x}\frac{1}{\sqrt{2\alpha ^{2}\pi}}\exp
\left[ -\frac{\left( \varepsilon _{fi}-\omega \right) ^{2}}{2\alpha
^{2}}\right]
\nonumber \\
&=&\frac{1}{\sqrt{2\alpha ^{2}\pi}}\int_{-\infty}^{\infty}dk_{x}\exp %
\left[ -\frac{\left( k_{x}^{2}+Q\left( \zeta _{k_{y}}\right) \right) ^{2}}{%
8\alpha ^{2}}\right]  \label{intkx} \\
&=&\frac{1}{\alpha \sqrt{2\pi}}\sqrt{\frac{E_{S}}{\sqrt{\omega _{c}}}%
\left( \zeta _{k_y}+\zeta _{\min}\right)}\exp \left[
-\frac{E_{S}^{2}\left(
\zeta _{k_y}+\zeta _{\min}\right) ^{2}}{4\omega _{c}\alpha ^{2}}\right] K_{%
\frac{1}{4}}\left[ \frac{E_{S}^{2}\left( \zeta_{k_y}+\zeta _{\min
}\right) ^{2}}{4\omega _{c}\alpha ^{2}}\right]  \nonumber
\end{eqnarray}
where we have used the following formula (cf.~Eq.~(3.323) on p.307
of Ref.~\cite{Gradsh}):
\[
\int_{0}^{\infty}dx\exp \left[ -\beta ^{2}x^{4}-2\gamma
^{2}x^{2}\right]
=2^{-3/2}\frac{\gamma}{\beta}e^{\gamma ^{4}/2\beta ^{2}}K_{\frac{1}{4}%
}\left( \gamma ^{4}/2\beta ^{2}\right),
\]%
which holds for $\left| \arg \beta \right| <\frac{\pi}{4}$ and
$\left| \arg \gamma \right| <\frac{\pi}{4}$, and where $K_{\nu}(z)$
is a modified Bessel function (cf.~p.375 of Ref.~\cite{Abra65}).
When $Q\left( \zeta _{k_{y}}\right) <0$, the integration in
Eq.~(\ref{intkx}) must be done numerically.

\subsubsection{Plane Wave Limit of the Cross Section}

In the plane wave limit, $\alpha \rightarrow 0$, the integration
over $k_{x}$ (making use of Eq.~(\ref{quasidelta2})) becomes
\begin{equation}
\int_{-\infty}^{\infty}dk_{x}\delta \left( \varepsilon _{fi}-\omega
\right)
=\int_{-\infty}^{\infty}dk_{x}\delta \left( \frac{1}{2}k_{x}^{2}+\frac{1%
}{2}Q\left( \zeta _{k_{y}}\right) \right).
\end{equation}%
This integral is non-zero only when $\varepsilon _{fi}-\omega
=\frac{1}{2}k_{x}^{2}+%
\frac{1}{2}Q\left( \zeta _{k_{y}}\right) =0$, i.e., when we have
strict energy conservation. For non-zero real $k_{x}$,   we should
thus require
\[
Q\left( \zeta _{k_{y}}\right) =\frac{2E_{S}}{\sqrt{\omega
_{c}}}\left( \zeta _{k_y}+\zeta _{\min}\right) < 0
\]%
or
\begin{equation}
\tilde{\zeta}_{k_{y}} \equiv - \zeta _{k_{y}} >  \zeta _{\min}.
\label{zetalimit}
\end{equation}
Thus we have that
\begin{eqnarray*}
&&\int_{-\infty}^{\infty}dk_{x}\delta \left( \frac{1}{2}k_{x}^{2}+\frac{1}{%
2}Q\left( \zeta _{k_{y}}\right) \right) \\
&=&\frac{1}{|Q|^{1/2}} \int_{-\infty}^{\infty}dk_{x} \left[ \delta
\left( k_{x} + |Q|^{1/2} \right) + \delta \left( k_{x} -
|Q|^{1/2}\right) \right] \\
&=&\frac{2\omega _{c}^{1/4}}{\sqrt{2E_{S}}}\frac{1}{\sqrt{-\zeta
_{k_{y}}-\zeta _{\min}}},
\end{eqnarray*}
In the plane wave limit, we have then (converting  $\zeta
_{k_{y}}$
   to $\tilde{\zeta}_{k_{y}}$)
\begin{equation}
\sigma _{\alpha= 0}^{(1)} =\frac{8\pi ^{2}C_{i}^{2}\omega
_{c}^{9/4}}{c\omega (\omega ^{2}-\omega
_{c}^{2})^{2}}\frac{1}{\sqrt{2E_{S}}}\sum\limits_{n_{z}=0}^{\infty
}\int_{\zeta _{\min}}^{\infty}\frac{d\tilde{\zeta}_{k_{y}}}{\sqrt{\tilde{%
\zeta}_{k_{y}}-\zeta _{\min}}}\left[ \tilde{\zeta}_{k_{y}}\omega
_{c}g_{n_{z}}\left( \tilde{\zeta}_{k_{y}}\right) +\omega
g_{n_{z}}^{\prime}\left( \tilde{\zeta}_{k_{y}}\right) \right] ^{2}.
\label{crs1stpln}
\end{equation}

We consider now two limiting cases, corresponding to weak static
magnetic and electric fields respectively.

\paragraph{The weak magnetic field limit.}
The plane wave cross section in Eq.~(\ref{crs1stpln}) can be
   simplified when the cyclotron frequency, $\omega _{c}$, is
much smaller than the laser frequency, $\omega$, i.e., $\omega
_{c}\ll \omega $. In this case,
\begin{equation}
\sigma _{\alpha =0,\omega _{c}\ll \omega}^{(1)} =
\frac{3\sigma ^{0}}{k^{3}}\frac{\omega _{c}^{9/4}}{\sqrt{2E_{S}}}%
\sum\limits_{n_{z}=0}^{\infty}\int_{\zeta _{\min}}^{\infty}d\tilde{\zeta}%
_{k_{y}}\frac{g_{n_{z}}^{\prime 2}\left( \tilde{\zeta}_{k_{y}}\right)}{%
\sqrt{\tilde{\zeta}_{k_{y}}-\zeta _{\min}}},  \label{crsdelos}
\end{equation}
where we have defined
\begin{equation}
\sigma ^{0} =\frac{8\pi ^{2}C_{i}^{2}}{3c\omega ^{3}}k^{3},
\label{crs_zerofld}
\end{equation}
in which $\sigma ^{0}$ is the photodetachment cross section for
H$^-$ in the monochromatic field limit in the absence of any
static fields, and  $k$ is the magnitude of the detached
electron's momentum, $k^{2}=2E_{f}=2\left( \omega +\varepsilon
_{i}\right)$.

We note that our weak magnetic field result in
Eq.~(\ref{crsdelos}) agrees with the formula of  Peters and Delos
(see Eqs.~(3.6) and~(3.7a) of Ref.~\cite{Peter93b}).
Eq.~(\ref{crsdelos})  agrees also with Fabrikant's result (see
Eq.~(53) of Ref.~\cite{Fabr91}) except for the extra term in his
formula that accounts for final-state interaction of the electron
with the atomic residue.

\paragraph{Weak static electric field  limit.}
In the limit  $E_{S}\rightarrow 0$, we have that $\ \zeta _{\min
}\rightarrow -\infty$. And we have also
\begin{equation}
   \lim_{E_{S}\rightarrow
0}\sqrt{2E_{S}}\sqrt{\tilde{\zeta}_{k_{y}}-\zeta _{\min}} =\omega
_{c}^{1/4}\sqrt{2\omega _{c}\left[ \left( \varepsilon _{i}+\omega
\right) /\omega _{c}-(n_{z}+\frac{1}{2})\right]}.
\end{equation}
Substituting this  result into Eq.~(\ref{crs1stpln}) and carrying
out the  integration involving  the Hermite polynomials, we obtain
\begin{equation}
\sigma _{\alpha =0,E_{S}=0}^{(1)}=\frac{8\pi ^{2}C_{i}^{2}\omega _{c}^{2}}{%
c\omega (\omega +\omega _{c})^{2}}\sum\limits_{n_{z}=0}^{n_{1}}\left[ \frac{%
\omega ^{2}+\omega _{c}^{2}}{(\omega -\omega _{c})^{2}}n_{z}+\frac{1}{2}%
\right] \frac{1}{\sqrt{2\omega _{c}\left[ \left( \varepsilon
_{i}+\omega \right) /\omega _{c}-(n_{z}+\frac{1}{2})\right]}}
\label{crs_gao}
\end{equation}
where the upper limit of summation, $n_1$, is the largest integer
that satisfies,
\[
n_{1}<\left[ \frac{\varepsilon _{i}+\omega}{\omega
_{c}}-\frac{1}{2}\right].
\]
   Eq.~(\ref{crs_gao}) is exactly the same as Gao's
result for the one-photon  detachment cross section in a static
uniform magnetic field (see Eq.~(31) of Ref.~\cite{Gao90a}).

\subsubsection{Cross Section for the Double Pulse Case}

From Eqs.~(\ref{crsdbl0}),  ~(\ref{transrate}),~(\ref{crsdef}) and
(\ref{flux2}), it is easy to show that for the double laser pulse
case, the cross section is given by
\begin{eqnarray}
\sigma _{\text{dbl}}^{(1)} &=&\frac{4\pi ^{2}C_{i}^{2}\omega _{c}^{2}}{%
c\omega (\omega ^{2}-\omega
_{c}^{2})^{2}}\sum\limits_{n_{z}=0}^{\infty}\int_{-\infty}^{\infty
}d\zeta _{k_{y}}\left[ \zeta _{y}\omega _{c}g_{n_{z}}\left( \zeta
_{k_{y}}\right) +\omega g_{n_{z}}^{\prime}\left(
\zeta _{k_{y}}\right) \right] ^{2}  \nonumber \\
&&\times \int_{-\infty}^{\infty}dk_{x}%
\frac{1+\cos \left[ \left( \varepsilon _{fi}-\omega \right) \tau
-\beta \right]}{1+\cos \beta \exp (-\alpha ^{2}\tau ^{2}/2)}
\overline{\delta}_{\alpha}\left( \varepsilon _{fi}-\omega \right).
\label{DblPulseCS}
\end{eqnarray}
We note that for $\tau =\beta =0$, this formula reduces to the
single pulse  result  in Eq.~(\ref{crs1st}), as it should  (cf.
Eq.~(\ref{laserpulsedbl})).

\section{Connections to Classical Closed Orbits}

In the previous section we have derived a general quantum mechanical
expression for the (effective) photodetachment cross section for
H$^{-}$ by a short laser pulse in the presence of crossed static
electric and magnetic fields. We  have also shown that our plane
wave limit result (given by Eq.~(\ref{crs1stpln})) reduces  for the
limiting cases of weak
 static magnetic (cf. Eq.~(\ref{crsdelos})) or weak static electric (cf.
Eq.~(\ref{crs_gao}))  fields to known results of others. Magnetic
field strengths, $B$,  that are readily available at present in the
laboratory are weak in the sense that they satisfy the relation,
$\omega _{c}\ll \omega $. Therefore the quantum result for the
photodetachment cross section in the plane wave limit given in
Eq.~(\ref{crsdelos}) is of great interest owing to the possibility
of experimental measurements with currently available technology. In
this section we analyze this equation for the purpose of making
connection with the classical closed orbits analyzed by Peters and
Delos~\cite{Peter93}. This connection will prove useful for
interpreting some of the numerical predictions presented in the next
section.

For $\omega\gg\omega_{c}$, the detached electron energy lies in
the region of large $n_{z}$.  In this limit the integrand in
Eq.~(\ref{crsdelos}) becomes highly oscillatory, as may be seen by
considering the large $n_{z}$ (Plancherel-Rotach) limit of the
Hermite function, $g_{n_{z}}$~\cite{Szego39}:
\begin{equation}
g_{n_{z}}(\tilde{\zeta}_{k_{y}})\simeq\sqrt{\frac{2}{\pi
\sqrt{2n_{z}(1-\eta ^{2})}}}\left\{\sin \left[ \left(
n_{z}+\frac{1}{2}\right) \left( \eta \sqrt{1-\eta ^{2}}-\arccos \eta
\right) +\frac{3\pi}{4}\right]+O(n_{z}^{-1})\right\},
\label{gnzzetay}
\end{equation}
where we have defined %%
\begin{equation}
\eta \equiv \tilde{\zeta}_{k_{y}}/\sqrt{2n_{z}+1}, \label{yita}
\end{equation}
and note that
\[
\epsilon \leqslant \arccos \eta\leqslant \pi -\epsilon
,\text{}\epsilon \rightarrow 0^{+},
\]
  {\it{i.e.}},  the argument of the Hermite function,
$g_{n_{z}}(\tilde{\zeta}_{k_{y}})$, must lie between the classical
turning points:
\[
  -\sqrt{2n_{z}+1}<\tilde{\zeta}_{k_{y}}<\sqrt{2n_{z}+1}.
\]
The function  $g_{n_{z}}^{\prime}\left( \tilde{\zeta}_{k_{y}}\right)
$ that occurs in the integrand of Eq.~(\ref{crsdelos}) may be
calculated by differentiation of Eq.~(\ref{gnzzetay}) with respect
to $\tilde{\zeta}_{k_{y}}$, as follows:
   \begin{eqnarray}
   g_{n_{z}}^{\prime}\left( \tilde{\zeta}_{k_{y}}\right)  &=&
\frac{\partial \eta}
   {\partial \tilde{\zeta}_{k_{y}}}
    \frac{\partial}{\partial \eta} g_{n_{z}}
   \left( \tilde{\zeta}_{k_{y}}\right)
   = \frac{1}{\sqrt{2 n_z +1}} \sqrt{\frac{2}{\pi \sqrt{2 n_z}}}
   \nonumber \\
   &\times& \left\{\frac{\eta}{2} \left(1- \eta^2 \right)^{-\frac{5}{4}}
   \sin\left[ S(n_z, \eta)\right]
   + (2n_z + 1) \left(1- \eta^2 \right)^{\frac{1}{4}}
   \cos\left[ S(n_z, \eta)\right]  \right\},
\label{gnderivappr1}
\end{eqnarray}
where we have defined the phase
\begin{equation}
S(n_{z},\eta )=\left( n_{z}+\frac{1}{2}\right) \left( \eta \sqrt{1-\eta ^{2}}%
-\arccos \eta \right) +\frac{3\pi}{4}.
  \label{sphase}
\end{equation}
    %% %%
Assuming  that $n_z\gg1$, Eq.~(\ref{gnderivappr1}) can be
simplified (in particular, the first term within the curly
brackets can be ignored in comparison with the second term), so
that we obtain:
\begin{equation}
g_{n_{z}}^{\prime}\left( \tilde{\zeta}_{k_{y}}\right) \simeq
\sqrt{\frac{2}{\pi}} (2n_z)^{\frac{1}{4}} (1-\eta^2)^{\frac{1}{4}}
\cos\left[ S(n_z, \eta)\right]. \label{gnderivappr2}
\end{equation}

Owing to the fact that $n_z$ is large, the phase function
$S(n_z,\eta)$ changes significantly as $\eta$ varies (cf.
Eq.~(\ref{sphase})), so that $g_{n_{z}}^{\prime}\left(
\tilde{\zeta}_{k_{y}}\right)$ oscillates rapidly as a function of
$\tilde{\zeta}_{k_{y}}$.  From Eq.~(\ref{crsdelos}) we see that the
magnitude of the photodetachment cross section will have the highest
maxima when the squares of the various $g_{n_{z}}^{\prime}\left(
\tilde{\zeta}_{k_{y}}\right)$ functions that are summed (over $n_z$)
have their maxima and minima in phase with each other, i.e., when
neighboring phase functions differ by an integer multiple of $\pi$:
\[
S\left(n_{z},\eta(n_z)\right
)-S\left(n_{z}-1,\eta(n_z-1)\right)\simeq\frac{d}{d
n_{z}}S(n_{z},\eta )=j\pi \text{,  where }j=0,\pm 1,\pm 2,...
\]
This condition is similar to that found for the largest local
maxima in the photodetachment cross section of $H^-$ in the
presence of parallel static electric and magnetic
fields~\cite{Wang97}.  We compute the total derivative of
$S(n_z,\eta)$ as: %%
\begin{equation}
\frac{d}{d n_{z}}S(n_{z},\eta )=\frac{\partial S}{\partial n_z}  +
\frac{\partial S}{\partial\eta} \frac{\partial \eta}{\partial
n_z}\label{totderivS}
\end{equation}
   The partial derivatives of the phase $S(n_z,\eta)$ follow
straightforwardly from the definition in Eq.~(\ref{sphase}).  The
partial derivative,  ${\partial \eta}/{\partial n_z}$, is calculated
using the definition in Eq.~(\ref{yita}) and the expression for
$\tilde{\zeta}_{k_{y}}$ obtained from the energy conservation
condition, $\varepsilon _{fi}-\omega=0$, together with
Eqs.~(\ref{finaleps})-(\ref{zetamin}) and Eq.~(\ref{zetalimit}).
After some straightforward algebra, one obtains:
\begin{equation}
\frac{d}{d n_{z}}S(n_{z},\eta )=-\arccos \eta +\sqrt{1-\eta
^{2}}\frac{\omega _{c}}{E_{S}}\sqrt{2\varepsilon _{n_{z}}}=j\pi.
\label{closeorb1}
\end{equation}

The condition~(\ref{closeorb1}) for the highest local maxima in
the photodetachment cross section~(\ref{crsdelos})  may be
re-written in terms of the scaled energy $\varepsilon$, defined by
\begin{equation}
\varepsilon =\varepsilon _{n_{z}}\left( \frac{\omega
_{c}}{E_{S}}\right) ^{2}, \label{scaledeps}
\end{equation}
and the angle $\varphi$, defined by
\begin{equation}
\varphi =\frac{\pi}{2}-\arccos \eta, \label{scaledphi}
\end{equation}
to obtain:
\begin{equation}
\cos \varphi +\frac{\varphi}{\sqrt{2\varepsilon
}}-\frac{1}{\sqrt{2\varepsilon}}%
\left( j+\frac{1}{2}\right) \pi =0.  \label{stationphase1}
\end{equation}
This result is identical to the classical equation expressing the
relationship of the azimuthal angle $\varphi$ and the scaled energy
$\varepsilon$ for a closed orbit of an electron in crossed fields (see
Eq.~(3.12) of Ref.~\cite{Peter93}).

The classical Hamiltonian corresponding to the quantum
Hamiltonian~(\ref{hamiltonianmom}) for the detached electron  is
given by
\begin{eqnarray}\label{classicalH}
     H_{\text{cls}} &=& \frac{1}{2} \omega _{c}^2 z ^2 +  \omega _{c}
z \left(p_y - \frac{E_{S}}{\omega _{c}} \right) + \frac{1}{2}p_z^2
+  \frac{1}{2}p_x^2
     + \frac{1}{2}p_y^2 \nonumber \\
     &=& \frac{1}{2}p_x^2 +  \frac{E_{S}}{\omega _{c}} p_y
      + \frac{1}{2}p_z^2 + \frac{1}{2} \left[\omega _{c} z +  \left(p_y
     - \frac{E_{S}}{\omega _{c}} \right)\right]^2
     - \frac{1}{2} \frac{E_{S}^2}{\omega _{c}^2}. \label{classH}
\end{eqnarray}
Denoting
\begin{equation}\label{circuez}
  \varepsilon_z = \frac{1}{2}p_z^2 + \frac{1}{2} \left[\omega _{c} z +
\left(p_y - \frac{E_{S}}{\omega _{c}} \right)\right]^2,
\end{equation}
and introducing the following scaled coordinate, momentum, and time
variables,
  \begin{equation}\label{scalmom1}
     \tilde{\mathbf{q}}  =  \frac{\omega_c^2}{E_S}\mathbf{q},
\end{equation}
\begin{equation}\label{scalmom2}
     \tilde{\mathbf{p}}  =  \frac{\omega_c}{E_S}\mathbf{p},
\end{equation}
\begin{equation}\label{scaltime}
     \tilde{t} =  \omega_c t,
\end{equation}
 Eq.~(\ref{classH}) may be rewritten  as
\begin{equation}
     E = \varepsilon + \frac{1}{2} \tilde{p}_x^2   + \tilde{p}_y -
\frac{1}{2},
  \label{scaledHcall}
\end{equation}
where $E = \omega _{c}^{2} (\omega + \varepsilon_i)/{E_{S}^{2}}$ is
the scaled total energy and $\varepsilon$ is given by
Eq.~(\ref{scaledeps}) in which the quantum energy
$\varepsilon_{n_z}$ (cf. Eq.~(\ref{epsnz})) is replaced by the
 classical energy $\varepsilon_z$ in Eq.~(\ref{circuez}).

With the help of Eqs.~(\ref{epsnz}), (\ref{zetaky1})
 and~(\ref{zetalimit}), it is easy to show from the definition~(\ref{yita}) that
\begin{equation}
       \eta = -\frac{p_y - E_S/\omega_c}{\sqrt{2
       \varepsilon_{n_z}}},
       \label{yita2}
\end{equation}
which can be rewritten, in terms of scaled energies, as
 \begin{equation}
\eta = \frac{\varepsilon - (E-1/2)}{\sqrt{2\varepsilon}},
       \label{yita3}
 \end{equation}
 by using the energy conservation equation (\ref{scaledHcall}) and the fact
 that $p_x = 0$ for  closed orbits.
Substituting Eq.~(\ref{yita3}) into Eq.~(\ref{scaledphi}),
 we can rewrite Eq.~(\ref{stationphase1}) as
\begin{equation}
  \Lambda(\varepsilon) \equiv \sqrt{2\varepsilon -\left[\varepsilon -(E-1/2)\right]^{2}}
 -\arccos\left[ \frac{\varepsilon -(E-1/2)}{\sqrt{2\varepsilon}}\right]
 =j\pi. \label{stationphase2}
\end{equation}
 For a given scaled total energy $E$, the number of the solutions
 of Eq.~(\ref{stationphase2}) gives the
 total number of closed
 orbits.   The return time of a closed orbit in  crossed
 fields is given by~\cite{Peter93}
\begin{equation}
T_{\text{ret}}= 2(\omega_c)^{-1} \sqrt{2\varepsilon}\cos \varphi,
 \label{Treturn}
\end{equation}
which can be rewritten with the help of Eqs.~(\ref{scaledphi}) and~(\ref{yita3}) as
\begin{equation}
T_{\text{ret}}= 2(\omega_c)^{-1} \sqrt{2\varepsilon - \left[
\varepsilon - (E-1/2)\right]^2}.
 \label{Treturn2}
\end{equation}

 As discussed in~\cite{Peter93}, there
 exists a very important group of closed orbits whose total
 energies are given approximately (in the large energy  limit) by
  \begin{equation}
  E_j^b \simeq \frac{\pi^2}{2}\frac{E_S^2}{\omega_c^2}\left[\left(j+\frac{1}{2}\right)^2
  - \frac{3}  {\pi^2}\right],
  \label{bndengy1}
  \end{equation}
where $j=1, 2, 3,...$.  These are called boundary energies, because
for each $j$ a new closed orbit appears at the energy given by
Eq.~(\ref{bndengy1}) and for higher total energies this newborn
closed orbit will split (or ``bifurcate'') into a pair of closed
orbits with two different energies and return times, given by
Eqs.~(\ref{stationphase2}) and~(\ref{Treturn2}).

 Actually, each boundary energy defines the onset of  large oscillations
 in the cross section. However,  the largest amplitude oscillation    in the cross
 section occurs at a slightly higher energy at which a   different type of
 closed orbit occurs that has   a truly
 circular motion  in the drift frame in the $y$-$z$ plane.
 The energy of this orbit may be obtained by setting  the  initial momentum along the
  $y$ axis equal to the drift velocity, {\it i.e.},
  \begin{equation}
   p_y^0 = \frac{E_S}{\omega_c}.
   \label{py0cond}
  \end{equation}
  From the energy conservation equation (\ref{scaledHcall}) and the fact that
   $p_x=0$ for a closed orbit, we have
  $\varepsilon = E-1/2$.  Substituting this result into
  Eq.~(\ref{stationphase2}) gives
\begin{equation}
 \sqrt{2\varepsilon}
 =(j + \frac{1}{2} )\pi,
 \label{stationphase3}
\end{equation}
which in  unscaled variables corresponds to a  total energy equal to
  \begin{equation}
 \overline{E}_j  =  \frac{\pi^2}{2}\frac{E_S^2}{\omega_c^2}
 \left[\left(j+\frac{1}{2}\right)^2
  + \frac{1}  {\pi^2}\right].
  \label{bndengy2}
  \end{equation}
   Comparing Eqs.~(\ref{bndengy1}) and~(\ref{bndengy2}), one sees that the energy
   difference between the boundary orbits and the orbits having $p_y^0$ equal
   to the drift velocity is $2{E_S^2}/{\omega_c^2}$, independent of the value of $j$.
   Boundary closed orbits satisfy
   $\partial \Lambda(\varepsilon) / \partial \varepsilon =0$, which
   gives the relationship  $ \varepsilon = E+\frac{3}{2}$ in the
   large energy limit.  Closed orbits for which Eq.~(\ref{py0cond})
   applies have $ \varepsilon = E-\frac{1}{2}$.  From Eqs.~(\ref{Treturn2}) and~(\ref{stationphase3}), we find for these latter orbits that
\begin{equation}
T_{\text{ret}}=  (j +\frac{1}{2})   T_B,
 \label{Treturn1}
\end{equation}
where $T_B= {2\pi} / {\omega_c} $ is the cyclotron period and $j$ is
a positive integer. This formula is very similar to that obtained
for the case of parallel  static electric and magnetic fields
~\cite{Pete94, Wang97}, in which the largest oscillation amplitude
of the cross section corresponds to classical orbits for which for
an electron is ejected along the static field direction and
reflected by the static electric field such that its return time
satisfies $ T_{\text{ret}} = j T_B $. In the parallel fields case,
the motion in the plane perpendicular to the magnetic field is
simply cyclotron motion with period $T_B$. Classical closed orbits
having a return time equal to an integer multiple of $T_B$ are
associated with  the largest oscillations in the cross
section~\cite{Wang97}.

 In the crossed fields case, however, the situation is much more complicated.
However, since $\varepsilon_z$ (cf. Eq.~(\ref{circuez})) is
conserved, when the detached electron has an initial momentum
$p_y^0$ given by Eq.~(\ref{py0cond}) (and an initial position
$z=0$), the initial momentum along the $z$ axis  takes its maximum
value. This implies that this particular closed orbit starts out (in
the drift frame) aligned with the laser polarization direction.
These circular orbits (in the drift frame), having energies given by
Eq.~(\ref{bndengy2}), are associated with the largest amplitude
oscillation of the cross section.

%%%%%%%%%%%%%%%%%%%%%%%%%%%%%%%%%%%
%
%
% Results and Discussion Section
%
%
%%%%%%%%%%%%%%%%%%%%%%%%%%%%%%%%%%%

\section{Results and Discussion}
\begin{figure}
    \includegraphics[width=12cm]{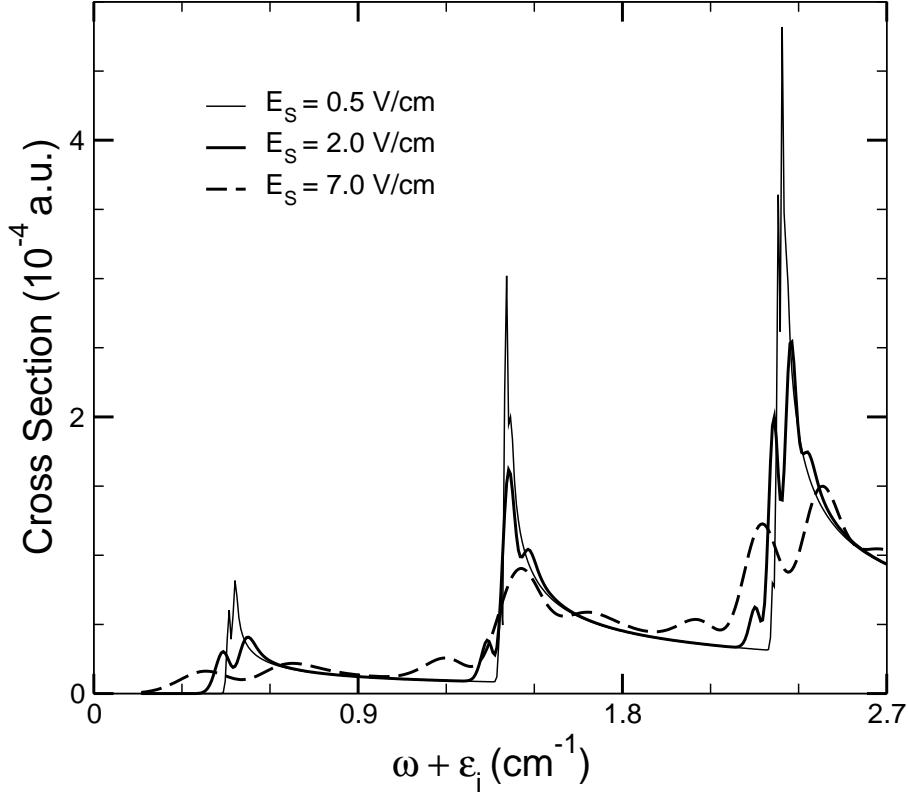}\\
    \caption{Photodetachment cross section for $H^-$ in the plane wave limit for a
static magnetic
    field $B=1$ T and three different values of the static electric field strength. Results are plotted versus detached electron kinetic energy above the zero field detachment threshold,
    $\omega + \varepsilon _{i}$, up to $2.7$ cm$^{-1}$.}
    \label{figure2}
\end{figure}
In this section we present numerical results based on the
quantum mechanical theoretical formulation presented above. We present first
plane wave limit results for the photodetachment cross section of $H^-$ in the presence of crossed static electric and magnetic fields over a much larger energy range than in prior works~\cite{Fabr91, Peter93, Peter93b}.  This large range allows us to demonstrate very clearly the signatures of the predicted classical closed orbits~\cite{Peter93}, both in the energy spectrum and in the time (i.e., Fourier transform) spectrum.  We
examine next the short laser pulse case, demonstrating first the
effects of laser pulse duration on the
photodetachment cross section.  We then examine the detached electron wave packet dynamics  in the $y$-$z$ plane and the possibility of modulating the detachment cross section by pump probe (Ramsey interference) techniques. The connection between the time development of the quantum wave packet of the detached electron and the predicted classical closed orbits is also discussed.

\subsection{Photodetachment Cross Section in the Plane Wave Limit}

\subsubsection{Static Electric Field Dependence for Near Threshold Energies}
\begin{figure}
\includegraphics[width=12cm]{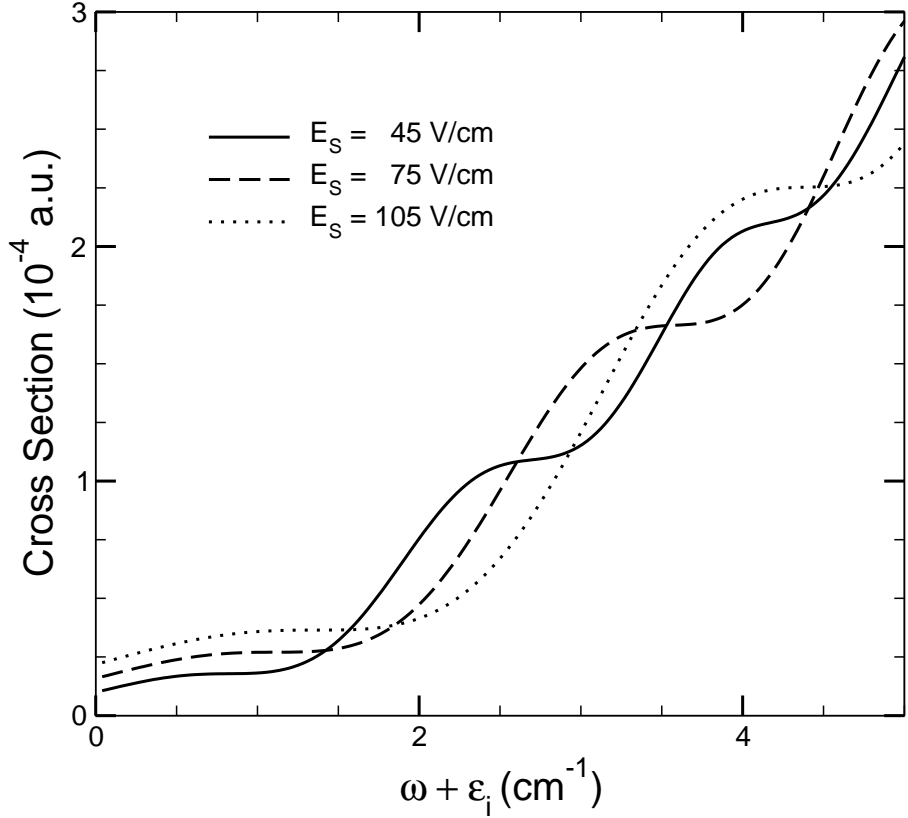}\\
\caption{Same as Fig.~\ref{figure2} but for three higher static
electric
     field strengths and detached electron kinetic energies up to $5$ cm$^{-1}$.}
     \label{figure3}
\end{figure}
   In Figs.~\ref{figure2} and~\ref{figure3} we present the photodetachment cross section for $H^-$ for a static magnetic field, $B = 1$ T,  and six different values of the static electric field, $E_S$.  Our quantum theory predictions are obtained from the plane wave limit result given in Eq.~(\ref{crs1stpln}).  One sees in Fig.~\ref{figure2} that, as noted by Fabrikant~\cite{Fabr91}, even a very small static electric field removes the known singularity in the detachment cross section for energies corresponding to integer multiples of the cyclotron frequency in the pure magnetic field case (see, e.g.,~\cite{Gao90a}).  In particular, for $E_S = 0.5$ V/cm, the behavior of the cross section is very similar to that of the pure
magnetic field case~\cite{Gao90a} (to which our results reduce in
the limit of zero static electric field, as shown in Sec. II.F.1(b)
above),  but without the cyclotron singularities.  On the other
hand, beginning with $E_S=7$ V/cm, the oscillatory modulation of the
cross section by the static electric field becomes obvious.  As
shown in Fig.~\ref{figure3} the frequency of this modulation
decreases as the static electric field magnitude increases, just as
is found for the case of a pure static electric field or for the
case of parallel static magnetic and electric fields (see,
e.g.,~\cite{Wang95}). One sees also that the cross section becomes
non-zero at the zero-field threshold owing to the lowering of the
threshold by the static electric field.
 \begin{figure}
\includegraphics[width=12cm]{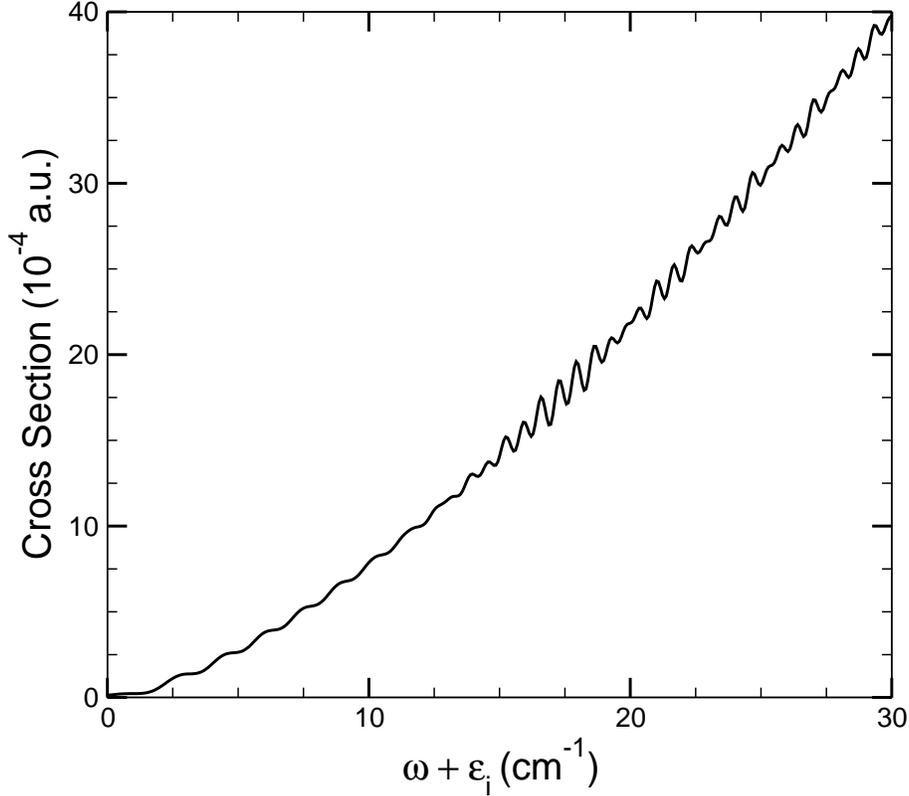}\\
\caption{Photodetachment cross section for $H^-$ in the plane wave
limit for $B=1$ T, $E_S=60$ V/cm,  and detached electron kinetic
energies up to $30$ cm$^{-1}$.}
     \label{figure4}
\end{figure}
For the present crossed static magnetic and electric field case, the
modulation of the cross section becomes increasingly complex the
higher the maximum total energy $E_f$ becomes. For a maximum
detached electron kinetic energy of 30 cm$^{-1}$, Fig.~\ref{figure4}
shows that the oscillatory modulations differ above and below
approximately 15 cm$^{-1}$.  For energies below 15 cm$^{-1}$, there
exists only a sinusoidal modulation. Above about 15 cm$^{-1}$,  the
modulation consists of more than one frequency and becomes more
complicated the higher in energy one looks.

In order to examine these structures in detail, it is instructive
to plot only the oscillatory part of the cross section, which is
defined by
\begin{equation}\label{crsosc}
      \sigma_{\text{osc}} =  \sigma _{\alpha \rightarrow 0}^{(1)} -
      \sigma^0,
\end{equation}
where $\sigma _{\alpha \rightarrow 0}^{(1)}$ is the total cross
section in the plane wave limit (given by Eq.~(\ref{crs1stpln})) and $\sigma^0$ is the photodetachment cross section in the absence of any external static fields (given by Eq.~(\ref{crs_zerofld})).
Figure~\ref{figure5} shows the oscillatory part of the cross
section over three different energy ranges,  corresponding to total energies  up to 60 cm$^{-1}$, 180 cm$^{-1}$, and 500 cm$^{-1}$ respectively.  While the oscillatory modulations of the cross section become increasingly dense and complex as the total energy increases, we see also that clear patterns in the spectra emerge and become more obvious the higher in energy we look.  The onset of these repetitive patterns is indicated in each panel by the vertical dotted lines, which represent the locations of the boundary energies~\cite{Peter93} defined by Eq.~(\ref{bndengy1}).  The peak amplitudes are indicated by the open triangles at the energies  defined by Eq.~(\ref{bndengy2}), which correspond to the locations of circular classical orbits in the drift frame.  The connection of these classical closed orbits and our quantum mechanical cross sections can be most easily investigated in the time domain, which we consider next.

\subsubsection{Fourier Transform Spectra and Closed Classical Orbits}
\begin{figure}
\includegraphics[width=12cm]{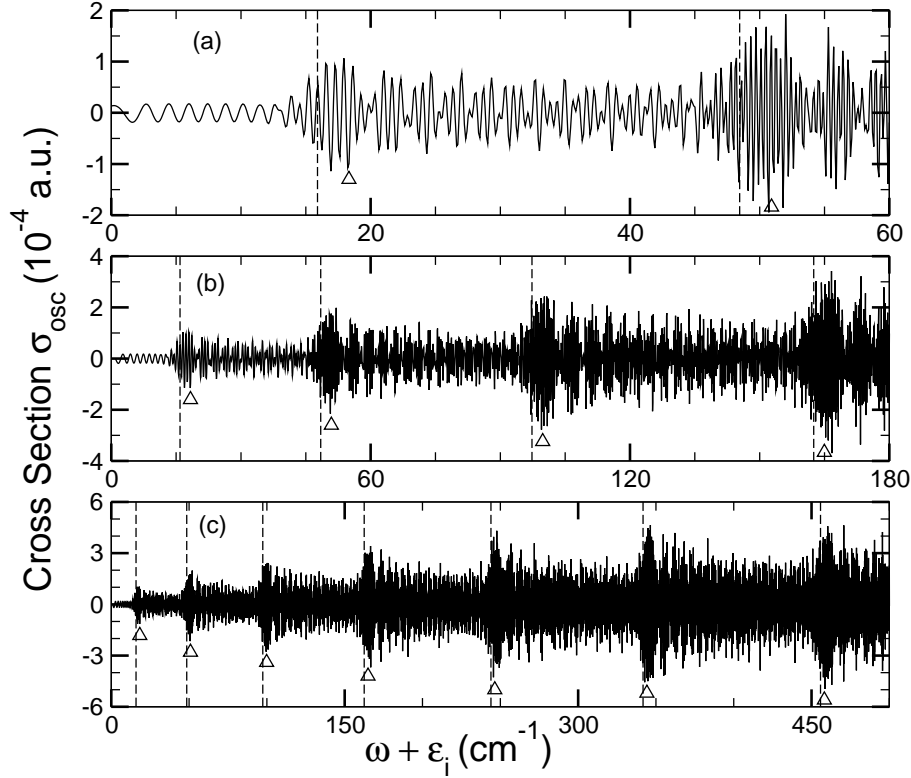}\\
\caption{The oscillatory part of the cross section, $\sigma_{\rm
{osc}}$ (cf. Eq.~(\ref{crsosc})),  for $B=1$ T and
     $E_S=60$ V/cm for $\omega+\varepsilon_i$ ranging from (a) 0 to 60 cm$^{-1}$; (b) 0 to 180 cm$^{-1}$; (c) 0 to 500 cm$^{-1}$. Dashed lines indicate the boundary energies (cf. Eq.~(\ref{bndengy1})) at which a new closed orbit appears~\cite{Peter93} and the open triangles indicate the energies at which the amplitude of the oscillatory part of the quantum cross section is expected to have a local maximum (cf. Eq.~(\ref{bndengy2})).}
     \label{figure5}
\end{figure}
\begin{figure}
    \includegraphics[width=12cm]{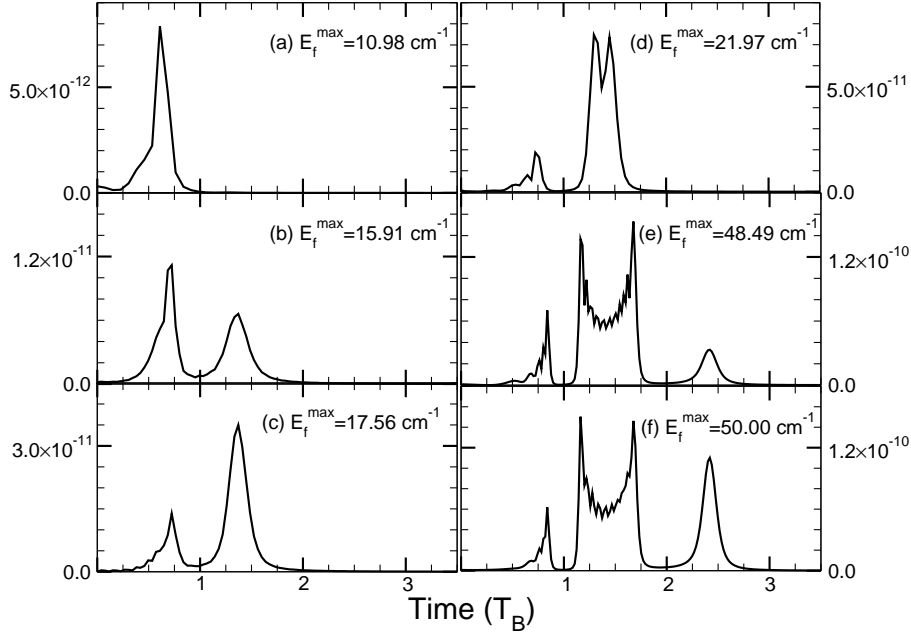}\\
    \caption{Fourier transform spectra for the
oscillatory part of the cross section (cf. Eq.~(\ref{crsosc}))
    for different maximum energies,  $E_f^{max} = \omega+\varepsilon_i$, as indicated in each panel.
   See text for a detailed description.}
    \label{figure6}
\end{figure}
The  Fourier transform of the oscillatory part of the
photodetachment cross section, $\sigma_{\text{osc}}$, is presented
in Fig.~\ref{figure6} for increasing values of the maximum total
detached electron kinetic energy, $E_f^{max}$. Times are given in
units of the cyclotron period, $T_B$. We see from
Fig.~\ref{figure6}(a) that at the lowest maximum energy,  only one
peak appears in the time spectrum. The peak position indicates the
return time (i.e., orbit period) of a closed classical orbit having
an energy of 10.98 cm$^{-1}$. As $E_f^{max}$ increases to 15.91
cm$^{-1}$ in (b), another peak emerges. This corresponds to the
first classical boundary energy (cf. Eq.~(\ref{bndengy1})) for $j =
1$. Unlike the case of classical dynamics, however,  where the
boundary energy is sharply defined, our calculations indicate that
the second peak in Fig.~\ref{figure6}(b) begins to appear around the
energy 13.5 cm$^{-1}$. Note also that the first peak in (b) shifts
to the right as compared to that in (a). This indicates that the
return time of the first closed orbit increases when the
   total energy increases. In panel (c) we observe that the second peak increases in magnitude while the first peak decreases in magnitude.  In panel (d), when the total energy equals  21.97 cm$^{-1}$, one observes  the bifurcation of the second peak that first appeared in panel (b).  For  $E_f^{max}=48.49$ cm$^{-1}$, in panel (e),
    one sees the appearance of a third peak around 2.4 $T_B$ and notice that the width of the splitting of the second peak increases
   as compared to that in (d). The left and
   right boundaries of the split and broadened second peak correspond to the
return times of the two
   bifurcated orbits at the maximum available total energy $E_f^{max}$.
  The serrated U-shaped region between the left
   and right boundaries of the split second peak correspond to the return times of the two bifurcated
   orbits for lower total energies (e.g., such as those two shown in panel (d) for a total energy of 21.97 cm$^{-1}$).  Finally, in panel (f) for a slightly higher total energy we observe that the third peak grows in magnitude relative to the first and second (split) peaks.

Consider now a much larger total final state energy, $E_f^{max}$ = 500 cm$^{-1}$.  The oscillatory part of the cross section is shown in Fig.~\ref{figure5}(c).  In this figure, the dashed lines correspond to the seven boundary orbit energies, given by Eq.~(\ref{bndengy1}), that appear for energies up to 500 cm$^{-1}$, and the triangles correspond to the seven circular orbit (in the drift frame) energies, given by Eq.~(\ref{bndengy2}).   The Fourier transform
spectrum for energies in the range from 0 - 500 cm$^{-1}$ is shown in Fig.~\ref{figure7}(a), while the Fourier transform of only the part of the spectrum in the range from 400 - 500 cm$^{-1}$ is shown in Fig.~\ref{figure7}(b). In (a), the open circles denote the return times (periods) of the 15 closed classical orbit solutions of Eq.~(\ref{stationphase2}) that exist for a total energy of  500 cm$^{-1}$.  These periods are calculated using Eq.~(\ref{Treturn2}) and the results are given in Table~\ref{table1} together with the corresponding orbit energies. (Note that for each $j>0$ and for a total energy $E$ not equal to one of the boundary energies given by Eq.~(\ref{bndengy1}), Eq.~(\ref{stationphase2}) has two solutions.) The open triangles, on the other hand, correspond to the circular orbits in the drift frame corresponding to the local maximum amplitudes of the oscillatory part of the cross section; their return times are given by the very simple Eq.~(\ref{Treturn1}).

\begin{table}[tbp]
\caption{Numerical solutions of Eq.~(\ref{stationphase2}) for the energies, $ \varepsilon_j^{\pm}$(in units of cm$^{-1}$), for the closed orbits that exist for  a total
energy $\omega+\varepsilon_i=500$ cm$^{-1}$ (302.869 in scaled
units) and the corresponding closed orbit periods, $ T_j^{\pm}$ (in units of $T_B$),
calculated from Eq.~(\ref{Treturn2}). Note that there is only one
solution for the case $j=0$.}
\begin{tabular}{|c|c|c|c|c|c|c|c|c|}
\hline
 $j$  & 0 &1&2&3&4&5&6&7 \\
 \hline
 \hline
 $ \varepsilon_j^-$ & 461.459& 462.389 &463.987&466.337&469.595&474.056&
 480.398& 491.231\\
 $ T_j^-$  & 0.959 &1.918 &2.876&3.831&4.783&5.730&6.665& 7.571 \\
 \hline
 $\varepsilon_j^+$ &---  &542.137 &541.0371&539.127&536.264&532.160&526.143&515.609 \\
 $ T_j^+$ &---&1.041 &2.082&3.126&4.172&5.224&6.286& 7.378\\
\hline
\end{tabular}%
\label{table1}%
\end{table}

Note that  the bowl-like structures appearing in Fig.
\ref{figure7}(a) above each open triangle result from the fact that closed orbits have different return times for each different total energy and from the fact that the Fourier transform spectrum in this figure results from a large range of total energies, i.e., from 0 to 500 cm$^{-1}$.  When we calculate the Fourier transform of the oscillatory part of the cross section over only the limited energy range, $400$ cm$^{-1}$ $\le  \omega+\varepsilon_i \le 500$ cm$^{-1}$, as in Fig.
\ref{figure7}(b), then we observe that the first 13 peaks are approximately located at the positions of the  first 13 closed classical orbit periods given in Table~\ref{table1}, which were calculated for a maximum total energy of $500$ cm$^{-1}$.  The energy region $0 \le  \omega+\varepsilon_i \le 400$ cm$^{-1}$ is thus inferred to be responsible for the bowl-like structures in Fig.
\ref{figure7}(a)  owing to the shifts of the lowest 13 classical orbit periods (having $j \le 6$) for lower total energies. The bowl-like structure remains between the 14th and 15th closed classical orbits (having $j = 7$) as this pair of orbits first occurs above a total energy of approximately 455 cm$^{-1}$ (cf. Fig.~\ref{figure5}(c)).   We also observe from the data in Figs.~\ref{figure5} and~\ref{figure7} that the
oscillation amplitude of the cross section becomes larger as the total energy increases.

\begin{figure}
    \caption{Fourier transform spectra of the oscillatory part of the cross section, $\sigma_{\rm
{osc}}$ (cf. Eq.~(\ref{crsosc})),  given in Fig.~\ref{figure5} (c)
calculated over two different energy ranges:  (a) $0 \le
\omega+\varepsilon_i \le 500$ cm$^{-1}$; (b) $400$ cm$^{-1}$ $\le
\omega+\varepsilon_i \le 500$ cm$^{-1}$. In (a) the open circles
indicate the return times (i.e., orbit periods) of the 15 closed
orbits for $\omega+\varepsilon_i=500$ cm$^{-1}$ (see
Table~\ref{table1}). Also in (a), the open triangles indicate the
return times of the circular closed orbits (in the drift frame)
having  $\tilde{p}_y^0=1$ (cf. Eqs.~(\ref{scalmom2})
and~(\ref{py0cond})); these return times are given by
Eq.~(\ref{Treturn1}).} \label{figure7}
\end{figure}
\subsection{Detachment by Short Laser Pulses}

Photodetachment by means of one or more short laser pulses differs
from that by a monochromatic laser. Most obviously,  the pulse
bandwidth affects the measured spectrum of detached electrons. In
addition, short laser pulses produce localized detached electron
wave packets whose motion in crossed fields can be investigated and
compared to classical predictions~\cite{Alber1991, Garraway1995,
Bluhm1996}.  Most interesting, perhaps, is the possibility of
controlling the modulation of the detachment spectrum by variation
of the parameters of one or more laser pulses. In the rest of this
section, we examine each of these topics in turn for the crossed
static electric and magnetic field case.

We note first, however, several previous works on related problems.
Ramsey interference effects resulting from photodetachment of H$^-$
by two short, coherent laser pulses as a function of their relative
phase was examined by Wang and Starace for the case of a static
electric field~\cite{Wang93} and the case of parallel static
electric and magnetic fields~\cite{Wang95}.  The latter
work~\cite{Wang95} showed that a large modulation of the effective
detachment probability can be achieved by optimizing the static
field magnitudes and the time delay between laser pulses, as
follows: the field magnitudes should be such that the classical time
for reflection of an electron back to the origin by the static
electric field equals an integer multiple of the harmonic oscillator
period for electron motion in the static magnetic field;  also, the
time delay of the second pulse should coincide with the classical
time for the electron's return to the origin.   For the case of a
single short laser pulse, Du~\cite{Du95} examined the
photodetachment of H$^-$ in the presence of a static electric field
using modified closed orbit formulas.  He showed that when the laser
pulse duration is shorter than particular closed orbit periods, then
those orbits no longer contribute to the photodetachment spectrum.
Finally, Zhao et al.~\cite{Zhao99} have derived a uniform
semiclassical formula for the photodetachment cross section of a
negative ion by a short laser pulse for the case of  parallel static
electric and magnetic fields.

 \begin{figure}
    \includegraphics[width=12cm]{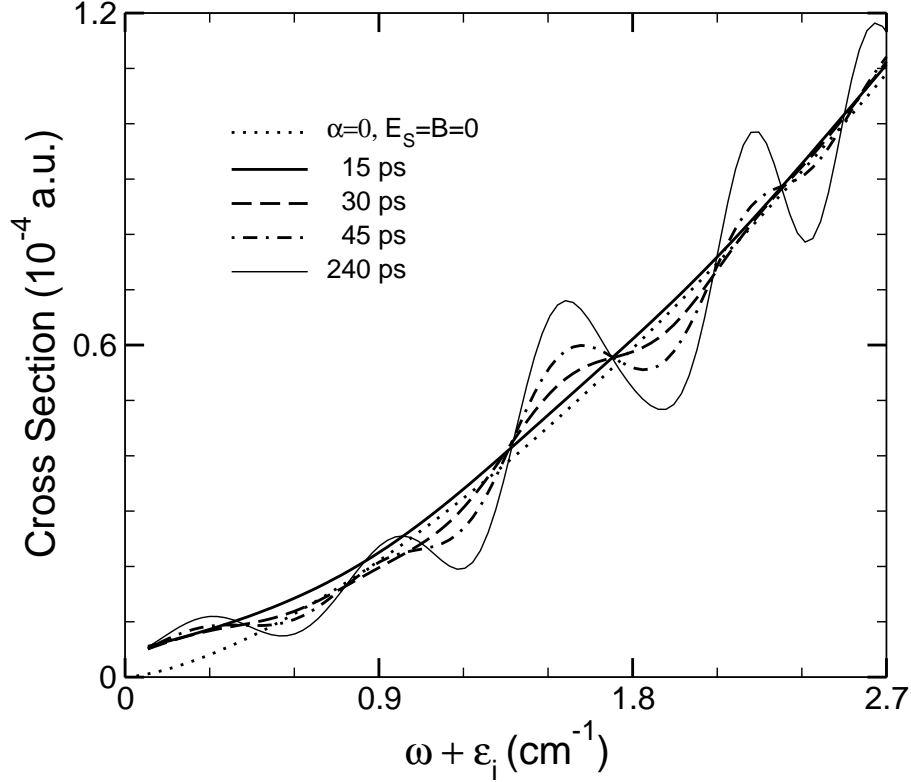}\\
    \caption{Effective photodetachment cross section of H$^-$ (cf. Eq.~(\ref{crs1st})) by  a single laser pulse of the form~(\ref{laserpulse}) with four different pulse durations (cf. Eq.~(\ref{Tpulse}) in the presence of crossed static electric and magnetic fields, $E_S = 15$ V/cm and $B = 1$ T. Results are plotted vs. electron kinetic energy beginning from the zero-field threshold. Also
shown (dotted line) is the photodetachment cross section for the
case of a continuous (monochromatic) laser without any external
static fields present.}
    \label{figure8}
   \end{figure}

\subsubsection{Pulse Duration Effects}

The fundamental difference between using a short laser pulse and
using a continuous (monochromatic) laser is the bandwidth of the
short laser pulse. In the former case, the laser pulse will excite a
group of final states that form an electron wave packet,  whereas in
the latter case only a well-defined final state will be reached.  In
our present case of detachment in the presence of crossed static
electric and magnetic fields, the spacing of Landau levels is very
small (0.93 cm$^{-1}$ for $B=1$ T), so that one expects that even a
quite long pulse having a duration of several picoseconds will have
considerable finite bandwidth effects on the detached electron wave
packet and its dynamics.

In Fig.~\ref{figure8}, we present the effective total cross section
(cf. Eq.~(\ref{crs1st})) for a laser pulse of the
form~(\ref{laserpulse}) (with $\tau = \beta = 0$) and four different
pulse durations (cf. Eq.~(\ref{Tpulse})) in the presence of crossed
static electric and magnetic fields of magnitudes $E_S = 15$ V/cm
and $B=1$ T. As we see from Fig.~\ref{figure8},  for the longest
pulse duration, 240 ps, the effective cross section is identical
with that for a (monochromatic) plane wave. As the pulse duration
decreases, the modulation of the cross section is suppressed,
beginning at the highest energies shown and progressing to
structures at lower energies. Thus, when the pulse duration is
reduced to 45 ps,  the modulation structure beyond the energy 2.3
cm$^{-1}$ is largely suppressed. As the pulse duration is further
reduced to 30 ps, even the modulation between 1 and 2 cm$^{-1}$
decreases in magnitude. At the shortest pulse duration, 15 ps, the
oscillatory structure completely disappears and the effective cross
section becomes a smooth curve passing through the oscillatory cross
sections for longer pulse durations.   In this case, the cross
section is nearly identical to the one for detachment by a
continuous (monochromatic) laser in the absence of any external
static fields, as shown by the dotted line in Fig.~\ref{figure8}.
The major difference between these cases occurs near the threshold:
our short pulse effective cross section is finite at the zero static
field threshold, while the monochromatic field cross section, in
accordance with Wigner's threshold law, is zero at threshold. This
difference is due to the lowering of the detachment threshold by the
static electric field.

It is interesting to relate these changes in the structure of the
effective detachment cross sections as a function of pulse duration
to the energy positions of the known classical closed orbits
~\cite{Peter93, Peter93b}.  For the maximum energy 2.7 cm$^{-1}$
considered here, there are three closed orbits available for the
static field parameters we employ.  These three orbits have return
times (periods) of 30.69, 40.94 and 60.72 ps. As the energy
decreases to 1.8 cm$^{-1}$, the return times decrease to 28.94,
43.32 and 56.19 ps respectively. As the energy is further decreased
to 1 cm$^{-1}$, there are only two closed orbits having return times
of 26.95 and 48.58 ps respectively.  We observe in
Fig.~\ref{figure8} that as the laser pulse durations become shorter
than the closed orbit periods, the structure of the effective cross
section is reduced.  In particular, for the shortest pulse duration,
15 ps, which is smaller than any closed orbit period, all structure
has disappeared.

\subsubsection{Wave  packet dynamics}
\begin{figure}
    \includegraphics[width=15cm]{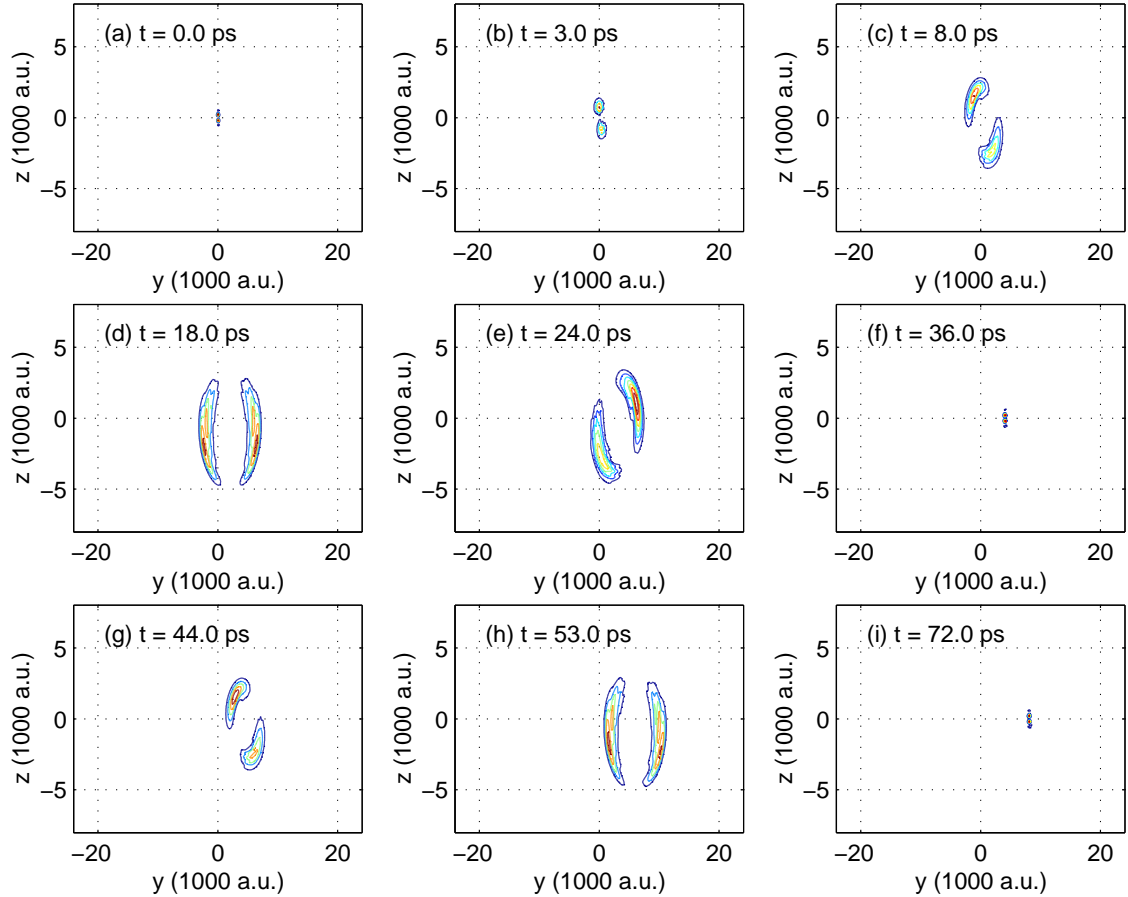}\\
    \caption{(color online). Contour plot snapshots of detached
electron wave  packet motion in the
    $y$-$z$ plane for the case of zero initial momentum in the $x$ direction
     [cf. Eq.~(\ref{wave  packetyxsgl})]. The static
    electric and magnetic field strengths are 60 V/cm and 1 T
    respectively. The laser pulse duration is $T_p=2$ ps and $t = 0$ corresponds to the end of this pulse. The total electron energy is $E_f=8$ cm$^{-1}$.}
    \label{figure9}
\end{figure}

We examine here the dynamics of a detached electron wave  packet produced by a short
laser pulse under the influence of crossed static electric and magnetic fields.
As discussed in Refs.~\cite{Peter93, Peter93b},  the oscillatory part of the
 photodetachment cross section produced by a monochromatic laser field may be
 associated with those closed orbits that exist for a given value of the photon
 energy. As discussed in the previous  section, the oscillatory part
 of the effective cross section produced by a short laser pulse is
 suppressed when the pulse duration is smaller than the classical
 orbit periods of those orbits that exist at the energy being considered.
 (See~\cite{Du95} for the related pure static electric field case.)
 From the correspondence between classical and quantum mechanics,
 we expect to observe that the    detached electron wave  packets
 produced by  short laser pulses will trace the paths of allowed
 classical closed orbits.

In order to illustrate the quantum wave packet motion corresponding
to the classical dynamics, we shall only consider  two-dimensional
quantum motion by imposing the restriction $p_x =0$. As discussed
above, the $x$ component of the momentum has to be zero in order for
there to be any closed orbits. In this case, the time-dependent
electron wave packet in coordinate space is given by
     \begin{equation}\label{wave  packetyxsgl}
      \psi_{\text{sgl}} ^{\text{wvpk}}(0,y,z,t) =
      \frac{1}{2\pi}\int_{-\infty}^{\infty}
      \psi_{\text{sgl}} ^{\text{wvpk}}(0,p_{y},z,t)
      e^{ip_y y},
      \end{equation}
   which is the Fourier transform of Eq.~(\ref{wavepacket1}),  taking
   $p_x=0$. A similar Fourier transform can be employed for the double
   pulse case in Eq.~(\ref{wvpkdble}).

   In Figs.~\ref{figure9}-\ref{figure11} we present snapshots of the detached electron
   wave packet for the case of a
   single laser pulse of the form~(\ref{laserpulse}) with pulse
   duration $T_p = 2$ ps. The time evolution starts from $t_0 = -T_p$. The static
    electric and magnetic field strengths are  taken to be  60 V/cm and 1 T
    respectively in all cases. Note that the cyclotron period is $T_B= 35.72$ ps
    for $B = 1$ T.
\begin{figure}
    \caption{(color online). Same as Fig.~\ref{figure9} but for a
total detached electron energy of $E_f=15.9$ cm$^{-1}$.}
    \label{figure10}
\end{figure}
 In Fig.~\ref{figure9}, we take the total energy $E_f$ to be 8
cm$^{-1}$. There is only one classical closed orbit  for this energy and these static field parameters. The  return time of the closed orbit is
calculated to be
    $T_{\rm{ret}}^{0} = 24.1$ ps (0.674 $T_B$). In Fig.~\ref{figure9}(a), we see that
    two electron wave packets are created at the peak intensity of the laser
    pulse on either side of the $z = 0$ axis, which correspond to electrons being ejected either along or opposite to the direction of increasing static electric field, $E_S$ (cf. Fig.~\ref{figure1}). After the end of the pulse in (b), the two electron wave packets move apart.   However, as time increases    we see in (c) that both wave packets are turned back by the
    external magnetic field. As shown in (c) and (d), each wave packet undergoes considerable spatial spreading.  The most interesting plot is shown in (e),
    where we see a large portion of the left hand wave packet
    sweep through the residual core (at the origin).  We note that the
    time corresponding to this snapshot is exactly the
    return time of the only classical closed orbit in the present case.
    It is return of this piece of the wave packet that leads to the
    regular sinusoidal  oscillation one sees in
Fig.~\ref{figure5}(a) below 15.9 cm$^{-1}$. As the time approaches 1 $T_B$ in (f),
    we see that the two wave packets refocus on the positive $y$  axis (the direction of drift motion) and that they pass through each other and continue their rotational motion during the
    next cyclotron period.  However, owing to the drift motion along the $y$-axis,
     we see in (h) that the left hand wave packet  is no longer able to return to the atomic core  in the second
    cyclotron cycle for this total energy. The two wave packets do  refocus further along the
    positive $y$-axis again at 2 $T_B$, as shown in (i).
\begin{figure}
    \caption{(color online). Same as Figs.~\ref{figure10}(e), (f), (g) and (h) except that the wave packet amplitudes are shown here in three-dimensions rather than as contour plots. Note the scale change in panel (b), which shows the re-focusing of the wave packet amplitude along the drift axis (i.e., away from the atomic core at the origin).}
    \label{figure11}
\end{figure}

    We note that the refocusing and the drift of the electron
    wave packets are exactly analogous to the classical dynamics  discussed  by Peters and Delos in~\cite{Peter93}. They showed that classical orbits with different initial
    conditions will refocus at various points along the drift axis. We note also that pump-probe experimental studies  of the related problem of the motion of pump-laser-produced Rydberg-state electron wave packets in the Rubidium atom in the presence of crossed fields have found enhancements of probe laser-produced ionization signals when the delay of the probe laser equals the orbital period of the appropriate closed classical orbit for this related problem~\cite{Yeaz93}.

 We present similar snapshots in time for the increased energy of 15.9
   cm$^{-1}$ in Fig.~\ref{figure10}.    At this energy, the first so-called boundary orbit~\cite{Peter93} may be populated (cf. Eq.~(\ref{bndengy1}) for $j = 1$).  There are thus two classical closed orbits that exist at this total energy, whose return times are  27.4  ps (0.766 $T_B$) and 48.8 ps (1.366 $T_B$). In Figs.~\ref{figure10}(e) and~\ref{figure10}(g) we see that different parts of the quantum electron wave packet return to the atomic core at these two times.  The most distinctive feature of the part of the electron wave packet that returns to the origin at approximately 49 ps (cf. Fig.~\ref{figure10}(g)) (which corresponds to the higher energy, classical boundary orbit) is that most of the arc in (g) passes through the atomic core at the origin (which we have  confirmed by observing the motion of the electron
    wave packet on a finer time scale). We note also that the energy of the classical boundary orbit corresponds to the abruptly increased amplitude of the oscillatory part of the cross section seen in Fig.~\ref{figure5}(a)
    around the energy location 15.9
   cm$^{-1}$ indicated by the first dashed line.

    The fact that electron wave packet amplitudes return to the region of the atomic
    core implies the possibility of modulating the detachment cross section,
    analogously  to
     the case of a monochromatic laser, as shown in Figs.~\ref{figure4}
    and~\ref{figure5}. However, the present wave packet studies show why for the
    crossed field case the modulation of the cross  section is  very small.
    Consider the three dimensional wave packet snapshots in Fig.~\ref{figure11},
    which are calculated for times corresponding to those in
    Figs.~\ref{figure10}(e), (f), (g), and (h). Owing to the spreading of the
    electron wave packet, when it returns to the origin the part of the wave packet
     that overlaps the origin is  nearly two orders of magnitude smaller than
     the probability in (f), in which the wave packet re-focuses along the
     drift axis, i.e., away from the origin. Because of such wave packet
     spreading and drift away from the origin, modulation of the photodetachment
     cross section in the crossed static field case  is necessarily small.
     For similar reasons, the use of short pulse, pump-probe type techniques
     to control the photodetachment cross section in the crossed static field
     case also results in only small modulations of the cross section,
     as we discuss next.

\subsubsection{Pump-Probe Coherent Control of the Effective Photodetachment Cross Section Using Short Laser Pulses}

The idea of using laser pulses shorter than electron wave
 packet orbital periods to control electron wave packet motion was initially
 formulated theoretically for Rydberg (i.e., bound) electron wave
 packets~\cite{ARZ1986}.  This idea was extended theoretically to photodetached
 (i.e., continuum) electron wave packets in the presence of external static fields,
 including static electric~\cite{Wang93} and parallel static electric and
 magnetic~\cite{Wang95} fields.  Experimentally, short pulse,
 pump-probe studies of photodetachment of O$^-$ in the presence of a static
 magnetic field demonstrated Ramsey interference between photodetached electron
 wave packets~\cite{Yuki97}.  Such Ramsey interference may also be demonstrated in
 the present crossed electric and magnetic field case.

\begin{figure}
    \includegraphics[width=12cm]{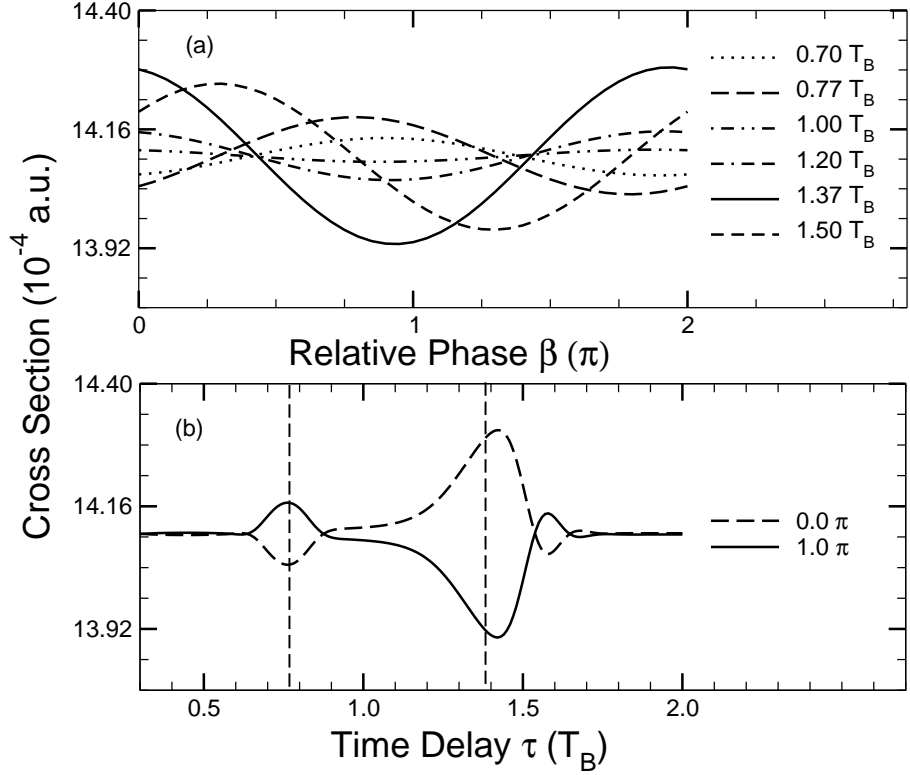}\\
    \caption{The effective cross section for  the double laser pulse
    case as modulated by: (a) the relative phase
    between the two pulses for several  time delays, as indicated;  (b) the time
    delay between the two pulses for two fixed relative phases, 0  and
    $\pi$. } \label{figure12}
\end{figure}
In Fig.~\ref{figure12} we present the effective total
photodetachment cross section (cf. Eq.~(\ref{DblPulseCS})) as a
function of the relative phase $\beta$ and the time
 delay $\tau$ between two laser pulses (cf. Eq~(\ref{laserpulsedbl})) for a total
 detached electron energy of 15.9 cm$^{-1}$ and for $E_S=60$ V/cm and $B=1$ T.
    The pulse duration $T_p$ of both pulses is taken to be 4 ps.
 Fig.~\ref{figure12}(a)  shows the
 dependence of the effective cross section on the relative phase, $\beta$, for six
 time delays, $\tau$, between the pulses.  One observes that the modulations of the
 cross section have local maxima for time delays of 0.77 $T_B$ ($\approxeq 27.5$ ps)
  and 1.37 $T_B$ ($\approxeq 49$ ps), which are precisely the return times of the two
  allowed classical closed orbits for a total electron energy of 15.9 cm$^{-1}$.
   However, the
modulation of the cross section for the larger time delay is much greater than for
the smaller time delay, which is consistent with the extent of electron wave packet
overlap with the origin shown in Figs.~\ref{figure10}(e) and (g). In other words,
in the latter case a large portion of the wave packet passes over the origin,
 which makes
the Ramsey interference with the newly produced wave packet amplitude
 (due to the second pulse) of greater amplitude.

Fig.~\ref{figure12}(b) shows the dependence of the effective cross
section on the time delay, $\tau$, for two relative phases, $\beta$,
between the pulses:   0 and 1 $\pi$. Fig.~\ref{figure12}(b) clearly
shows that the maxima and minima in the effective cross section as a
function of the time delay between the pulses occur for time delays
of 0.77 $T_B$ ($\approxeq 27.5$ ps) and 1.37 $T_B$ ($\approxeq 49$
ps), which are the orbital periods of the two allowed classical
closed orbits.  We see once again that the modulation of the
effective cross section is much larger for the classical closed
orbit having the larger time delay, as explained above.

%%%%%%%%%%%%%%%%%%%%%%%%%%%%%%%%%%%
%
%
% Conclusions Section
%
%
%%%%%%%%%%%%%%%%%%%%%%%%%%%%%%%%%%%

\section{Conclusions}

In this paper we have presented a detailed quantum mechanical analysis of detachment of a weakly bound electron by a short laser pulse in the presence of crossed static electric and magnetic fields.  For specificity, we have chosen the parameters of the initial state of the weakly bound electron as those appropriate for the outer electron in H$^-$.  In particular we have presented an analytic expression for the final state electron wave function, i.e., the wave function for an electron moving in the field of a laser pulse of arbitrary intensity as well as in crossed static electric and magnetic fields of arbitrary strengths.  The general detachment probability formulas we present may therefore be used to analyze multiphoton detachment in crossed fields (although we have not presented this analysis here, but instead have focused on the weak laser field case).

Based upon our analytic results for the detachment probability by a short laser pulse, we have defined an effective detachment cross section for the short pulse case that is shown to reduce, in the long pulse limit to results of others for the monochromatic, plane wave case.  Our effective cross section formula allows us to demonstrate the effects of the laser pulse duration, such as, e.g.,
that for pulse durations shorter than the period of a particular classical closed orbit, the features of that closed orbit in the photodetachment spectrum (for the plane wave case) will simply vanish.
By means of a stationary phase analysis,  we have derived a condition for the existence of closed classical orbits that agrees exactly with that  obtained by Peters and
Delos by a purely classical analysis~\cite{Peter93}. We have also illustrated the
bifurcation of the closed classical orbits  at the so-called boundary energies~\cite{Peter93} by
Fourier transforming the oscillatory part of our quantum cross
section (in the long pulse limit) over various ranges of the final state electron energy.

Finally, our analysis of the motion of detached electron wave
packets produced by a short laser pulse provides  a direct
comparison of quantum and classical features for the crossed static
electric and magnetic field problem.  We find that the dynamics of
our two-dimensional detached electron wave packets are  consistent
with the predictions of closed classical orbit
theory~\cite{Peter93}.
 We have also shown that wave packet spreading and the fact that wave
 packet refocusing only occurs at the origin in the drift frame means
 that control of electron detachment in crossed static fields by means
 of laser pulses is less effective than in the parallel static electric
 and magnetic field case~\cite{Wang95}.

\appendix

\section{\label{appendixA} Analytical Wave Function for Free Electron Motion in
 a Laser Field and
 Crossed Static Electric and Magnetic Fields}

 In this appendix we give the details of the solution of
 Eq.~(\ref{tdsemom}), which describes  free electron motion
 in a  laser field in the presence of crossed static electric
 and magnetic fields. The configuration
 of the external fields is shown in Fig.~\ref{figure1}.
 The solution  is clearly separable in momentum
space and has the following form:
\begin{eqnarray}
\psi _{f}^{(p)}(\mathbf{p},t) &=&\psi _{x}(p_{x},t)\psi
_{y}(p_{y},t)\psi _{z}(p_{z},t)\exp \left[
-i\frac{1}{2c^{2}}\int_{-\infty
}^{t}A_{L}^{2}(t^{\prime})dt^{\prime}\right]  \nonumber \\
&=&\delta (p_{x}-k_{x})\delta (p_{y}-k_{y})\psi _{z}(p_{z},t)  \nonumber \\
&&\times \exp \left[ -i\frac{1}{2}(k_{x}^{2}+k_{y}^{2})t-i\frac{1}{2c^{2}}%
\int_{-\infty}^{t}A_{L}^{2}(t^{\prime})dt^{\prime}\right],
\label{ansatz1AP}
\end{eqnarray}
where the $z$ component of the final state wave function satisfies
the following equation:
\begin{equation}
i\frac{\partial}{\partial t}\psi _{z}(p_{z},t)=\left[
-\frac{1}{2}\omega
_{c}^{2}\frac{\partial ^{2}}{\partial p_{z}^{2}}-i\omega _{c}\left( p_{y}-%
\frac{E_{S}}{\omega _{c}}\right) \frac{\partial}{\partial p_{z}}+\frac{1}{2}%
p_{z}^{2}+\frac{1}{c}p_{z}A_{L}(t)\right] \psi _{z}(p_{z},t).
\label{tdsepsizAP}
\end{equation}

In order to solve  Eq.~(\ref{tdsepsizAP}), we make the following
substitution:
\begin{equation}
\psi _{z}(p_{z},t)=\exp \left[ -i\frac{1}{\omega _{c}}(k_{y}-\frac{E_{S}}{%
\omega _{c}})p_{z}\right] \psi _{z1}(p_{z},t),  \label{solutpsizAP}
\end{equation}
which serves to eliminate the term involving the first derivative of
$p_{z}$ from the equation satisfied by $\psi_{z1}(p_{z},t)$,
\begin{equation}
i\frac{\partial}{\partial t}\psi _{z1}(p_{z},t)=\left[
-\frac{1}{2}\omega
_{c}^{2}\frac{\partial ^{2}}{\partial p_{z}^{2}}+\frac{1}{2}p_{z}^{2}+\frac{1%
}{c}p_{z}A_{L}(t)-\frac{1}{2}\left( k_{y}-\frac{E_{S}}{\omega
_{c}}\right) ^{2}\right] \psi _{z1}(p_{z},t).  \label{tdsepsiz1AP}
\end{equation}
Upon making the substitution,
\begin{equation}
\psi _{z1}(p_{z},t)=\exp \left[ i\frac{1}{2}\left(
k_{y}-\frac{E_{S}}{\omega _{c}}\right) ^{2}t\right] \psi
_{z2}(p_{z},t),  \label{solutpsiz1AP}
\end{equation}
Eq.~(\ref{tdsepsiz1AP}) gives the following equation for $\psi
_{z2}(p_{z},t)$:
\begin{equation}
i\frac{\partial}{\partial t}\psi _{z2}(p_{z},t)=\left[
-\frac{1}{2}\omega
_{c}^{2}\frac{\partial ^{2}}{\partial p_{z}^{2}}+\frac{1}{2}p_{z}^{2}+\frac{1%
}{c}A_{L}(t)p_{z}\right] \psi _{z2}(p_{z},t).  \label{tdsepsiz2AP}
\end{equation}

Eq.~(\ref{tdsepsiz2AP}) has the form of the equation for a forced
harmonic oscillator, which can be solved exactly~\cite{Husi53}: %%
\begin{equation}
\psi _{z2}(p_{z},t)= \exp \left[ -i\varepsilon _{n_{z}}t
+i\frac{\dot{\xi}(t)}{\sqrt{\omega _{c}^{3}}}\sqrt{2}\zeta
_{p_{z}}+i\int_{-\infty}^{t}L(t^{\prime})dt^{\prime}\right] \omega
_{c}^{-1/4}g_{n_{z}}\left( \sqrt{2}\zeta _{p_{z}}\right),
\label{solutpsipz2AP}
\end{equation}
where $\zeta _{p_{z}}$ and $\varepsilon _{n_{z}}$  are  given by
Eqs.~(\ref{zetapz}) and~(\ref{epsnz}) respectively.   The function
$\xi(t)$ is related to the vector potential of the laser pulse and
satisfies the differential equation given in Eq.~(\ref{odeksi}). The
function $L(t)$  is defined in Eq.~(\ref{tdsepsiz3b}) in terms of
$\xi (t)$. In Eq.~(\ref{solutpsipz2AP}),  the function
$g_{n_{z}}\left( x\right)$ is defined by
 Eq.~(\ref{gnzetay}).
Note that $g_{n_z}(x)$ is normalized, {\it{i.e.}},
\begin{equation}
\int_{-\infty}^{\infty}[g_{n_z}]^2\,dx = 1. \label{gnormAP}
\end{equation}
We note also that $g_{n_z}(x)$ is simply proportional to a parabolic
cylinder function (see Eq.~19.13.1 of Ref.~\cite{Abra65}):
\begin{equation}
g_{n_z}(x) = \frac{1}{\sqrt{n_z!\sqrt{\pi}}} U\left(
-n_z-\frac{1}{2} \sqrt{2}x\right).
   \label{gnvsUnAP}
\end{equation}

Combining  Eqs.~(\ref{solutpsizAP}), ~(\ref{solutpsiz1AP}),
and~(\ref{solutpsipz2AP}), we find for the solution of
Eq.~(\ref{tdsepsizAP}),
\begin{eqnarray}
\psi _{z}(p_{z},t) &=&\exp \left[ -i b(k_{y},t)\sqrt{2}\zeta
_{p_{z}}\right] \omega _{c}^{-1/4}  g_{n_{z}}\left(\sqrt{2}\zeta
_{p_{z}}\right)
    \nonumber \\
&&\times \exp \left[ -i\varepsilon _{n_{z}}t-i\frac{1}{\sqrt{\omega
_{c}}}\zeta _{k_{y}}\xi(t) +i\frac{1}{2}\omega _{c}\zeta
_{k_{y}}^{2}t+i\int_{-\infty}^{t}L(t^{\prime})dt^{\prime}\right],
\label{finalsoluzAP}
\end{eqnarray}
where  $b(k_{y},t)$ and $\zeta _{k_{y}}$   are given by
Eq.~(\ref{bt}) and~(\ref{zetaky1})  respectively. Finally,
substituting Eq.~(\ref{finalsoluzAP}) into Eq.~(\ref{ansatz1AP})
gives us the analytical expression for the final state wave
 function in momentum space given in Eq.~(\ref{finalwvmom}).

\section{\label{appendixB} Solution of Eq.~(\ref{odeksi}) for
$\xi (t)$}

In this appendix we present the solution for  the
laser-field-dependent function $\xi (t)$, which satisfies the
differential equation (\ref{odeksi}). The corresponding  Green's
function  satisfies,
\begin{equation}
\ddot{G}(t,t^{\prime})+\omega _{c}^{2}G(t,t^{\prime})=\delta
(t-t^{\prime}),  \label{greenAP}
\end{equation}
whose solution is:
\begin{equation}
G(t,t^{\prime})=\frac{1}{2\pi}\int_{-\infty}^{\infty
}\frac{e^{i\omega ^{\prime}(t-t^{\prime})}}{\omega _{c}^{2}-\omega
^{\prime 2}}d\omega ^{\prime}.
\end{equation}
In terms of this Green's function,  the solution of
Eq.~(\ref{odeksi}) is:
\begin{eqnarray}
\xi (t) &=&\int_{-\infty}^{\infty}G(t,t^{\prime})\left[
-\frac{\omega
_{c}^{2}}{c}A_{L}(t^{\prime})\right] dt^{\prime}  \nonumber \\
&=&\frac{\omega _{c}^{2}}{2\pi}\int_{-\infty}^{\infty}d\omega ^{\prime}%
\frac{e^{i\omega ^{\prime}t}}{\omega _{c}^{2}-\omega ^{\prime 2}}%
\int_{-\infty}^{\infty}dt^{\prime}\left[ -\frac{1}{c}A_{L}(t^{\prime})%
\right] e^{-i\omega ^{\prime}t^{\prime}}.  \label{ksait1AP}
\end{eqnarray}
Using the definitions in  Eqs.~(\ref{laserpulse})
and~(\ref{vectpot}), the integral over $t^{\prime}$ can be evaluated
(using integration by parts) to obtain,
\begin{eqnarray}
I(\omega^{\prime}) &=&\int_{-\infty}^{\infty}dt^{\prime}\left[
-\frac{1
}{c}A_{L}(t^{\prime})\right] e^{-i\omega^{\prime}t^{\prime}}  \nonumber \\
&=&\frac{\sqrt{\pi}}{i\omega^{\prime}}\frac{E_{0}}{2i\alpha}W(\omega
^{\prime}),  \label{IomegaprAP}
\end{eqnarray}
in which the function $W(\omega ^{\prime})$ is defined by %%
\[
W(\omega ^{\prime})=e^{-\frac{(\omega -\omega ^{\prime})^{2}}{4\alpha ^{2}}%
+i(\omega -\omega ^{\prime})\tau +i\beta}-e^{-\frac{(\omega +\omega
^{\prime})^{2}}{4\alpha ^{2}}-i(\omega +\omega ^{\prime})\tau
-i\beta},
\]
and where it has been assumed that $A_{L}(-\infty )=A_{L}(\infty
)=0$.

Thus $\xi (t)$ is now given by
\begin{eqnarray}
\xi (t) &=&\frac{\omega _{c}^{2}}{2\pi}\int_{-\infty}^{\infty
}d\omega ^{\prime}\frac{e^{i\omega ^{\prime}t}}{\omega
_{c}^{2}-\omega
^{\prime 2}}\frac{\sqrt{\pi}}{%
i\omega ^{\prime}}\frac{E_{0}}{2i\alpha}W(\omega ^{\prime})  \nonumber \\
&=&\frac{E_{0}}{2\alpha i}\frac{\sqrt{\pi}}{2}\left[
I_{0}(t)+I_{1}(t)+I_{2}(t)\right],  \label{ksait2AP}
\end{eqnarray}
where we have defined the following three integrals over $\omega
^{\prime}$:
\begin{eqnarray}
I_{0}(t) &=&\frac{1}{\pi i}\int_{-\infty}^{\infty}d\omega ^{\prime}\frac{1%
}{\omega ^{\prime}}W(\omega ^{\prime})e^{i\omega ^{\prime}t},
\label{Int0tAP} \\
I_{1}(t) &=&-\frac{1}{2\pi i}\int_{-\infty}^{\infty}d\omega ^{\prime}%
\frac{1}{\omega ^{\prime}-\omega _{c}}W(\omega ^{\prime})e^{i\omega
^{\prime}t},  \label{Int1tAP} \\
I_{2}(t) &=&-\frac{1}{2\pi i}\int_{-\infty}^{\infty}d\omega ^{\prime}%
\frac{1}{\omega ^{\prime}+\omega _{c}}W(\omega ^{\prime})e^{i\omega
^{\prime}t}.  \label{Int2tAP}
\end{eqnarray}

\subsection{Calculation of $I_{0}(t)$, $I_{1}(t)$ and $%
I_{2}(t)$}

Consider first $I_{1}(t)$, which satisfies the following relation:
\begin{eqnarray*}
\frac{d}{dt}[e^{-i\omega _{c}t} I_{1}(t)] &=&-\frac{1}{2\pi
}\int_{-\infty}^{\infty}d\omega ^{\prime}W(\omega ^{\prime
})e^{i\left( \omega ^{\prime}-\omega
_{c}\right) t} \\
&=&-\frac{i2\alpha}{\sqrt{\pi}}e^{-\alpha ^{2}(t-\tau )^{2}-i\omega
_{c}t}\sin (\omega t+\beta ).
\end{eqnarray*}
Thus, we have that
\begin{eqnarray}
I_{1}(t)&=&-\frac{i2\alpha}{\sqrt{\pi}}e^{i\omega
_{c}t}\int_{-\infty}^{t}e^{-\alpha ^{2}(t^{\prime}-\tau
)^{2}-i\omega _{c}t^{\prime}}\sin
(\omega t^{\prime}+\beta )dt^{\prime}  \label{I1tvalu1AP} \\
&=&-\frac{1}{2}e^{i\omega _{c}(t-\tau)}\left[
%TCIMACRO{\func{erf}}%
%BeginExpansion
\mathop{\rm erf}%
%EndExpansion
\left( \alpha (t-\tau )-\frac{i(\omega -\omega _{c})}{2\alpha}\right) +1%
\right] e^{-\frac{(\omega -\omega _{c})^{2}}{4\alpha ^{2}}+i\beta
+i\omega
\tau}  \nonumber \\
&&+\frac{1}{2}e^{i\omega _{c}(t-\tau)}\left[
%TCIMACRO{\func{erf}}%
%BeginExpansion
\mathop{\rm erf}%
%EndExpansion
\left( \alpha (t-\tau )+\frac{i(\omega +\omega _{c})}{2\alpha}\right) +1%
\right] e^{-\frac{(\omega +\omega _{c})^{2}}{4\alpha ^{2}}-i\beta
-i\omega \tau}.  \label{I1tvaluAP}
\end{eqnarray}%
Replacing $\omega _{c}$ by $-\omega _{c}$ in the above formula, we
get
\begin{eqnarray}
I_{2}(t) &=&-\frac{1}{2}e^{-i\omega _{c}(t-\tau)}\left[
%TCIMACRO{\func{erf}}%
%BeginExpansion
\mathop{\rm erf}%
%EndExpansion
\left( \alpha (t-\tau )-\frac{i(\omega +\omega _{c})}{2\alpha}\right) +1%
\right] e^{-\frac{(\omega +\omega _{c})^{2}}{4\alpha ^{2}}+i\beta
+i\omega
\tau}  \nonumber \\
&&+\frac{1}{2}e^{-i\omega _{c}(t-\tau)}\left[
%TCIMACRO{\func{erf}}%
%BeginExpansion
\mathop{\rm erf}%
%EndExpansion
\left( \alpha (t-\tau )+\frac{i(\omega -\omega _{c})}{2\alpha}\right) +1%
\right] e^{-\frac{(\omega -\omega _{c})^{2}}{4\alpha ^{2}}-i\beta
-i\omega \tau}.  \label{I2tvaluAP}
\end{eqnarray}
Setting $\omega _{c}$ equal to zero in Eq.~(\ref{I1tvaluAP}) and
multiplying by a factor of $-2$, we get
\begin{eqnarray}
I_{0}(t) &=&-\left[
%TCIMACRO{\func{erf}}%
%BeginExpansion
\mathop{\rm erf}%
%EndExpansion
\left( \alpha (t-\tau )+\frac{i\omega}{2\alpha}\right) +1\right] e^{-\frac{%
\omega ^{2}}{4\alpha ^{2}}-i\beta -i\omega \tau}  \nonumber \\
&&+\left[
%TCIMACRO{\func{erf}}%
%BeginExpansion
\mathop{\rm erf}%
%EndExpansion
\left( \alpha (t-\tau )-\frac{i\omega}{2\alpha}\right) +1\right] e^{-\frac{%
\omega ^{2}}{4\alpha ^{2}}+i\beta +i\omega \tau}. \label{I0tvaluAP}
\end{eqnarray}

\subsection{Some Relations for $\protect\xi (t)$ and Its
Derivatives}

   From  Eqs.~(\ref{ksait2AP}) and~(\ref{I1tvalu1AP}) (and equations similar
to~(\ref{I1tvalu1AP}) for $I_0(t)$ and $I_2(t)$), we have that
\begin{eqnarray}
\dot{\xi}(t) &=& \frac{E_{0} \sqrt{\pi}}{4 \alpha i} \left[
\dot{I}_{0}(t)+\dot{I}_{1}(t)+\dot{I}_{2}(t) \right] \\
&=&\frac{E_{0}\, \omega_{c} \sqrt{\pi}}{4 \alpha} \left[
I_{1}(t)-I_{2}(t) \right].  \label{ksitdotAP}
\end{eqnarray}
The second derivative of $\xi (t)$ is given by
\begin{eqnarray}
\ddot{\xi}(t) &=&\frac{E_{0}\,  \omega_{c} \sqrt{\pi
}}{4\alpha}\left[\dot{I}_{1}(t)-\dot{I}_{2}(t)\right]  \nonumber \\
&=&i\frac{E_{0}\, (\omega _{c})^{2} \sqrt{\pi}}{4\alpha}\left[
I_{1}(t)+I_{2}(t)\right].  \label{ksitdot2AP}
\end{eqnarray}
Writing now
   \begin{equation}
   \xi(t) = \sum_{i=0}^{2} \xi_i(t),
\label{ksisumAP}
   \end{equation}
   where
\begin{equation}
\xi _{i}(t)=\frac{E_{0}\sqrt{\pi}}{4\alpha i} I_{i}(t),
\label{ksiiAP}
\end{equation}
    Eqs.~(\ref{ksitdotAP}) and~(\ref{ksitdot2AP}) give
\begin{equation}
\dot{\xi} (t)= i\omega_c \left[\xi _{1}(t)-\xi _{2}(t) \right],
\label{ksait02AP}
\end{equation}%
\begin{equation}
\ddot{\xi} (t) = -\omega_c^2 \left[\xi _{1}(t)+\xi _{2}(t)\right].
\label{ksait03AP}
\end{equation}%
Substituting these equations into Eq.~(\ref{odeksi}) gives:
\begin{equation}
\xi _{0}(t)=-\frac{1}{c}%
A_{L}(t).  \label{ksi01AP}
\end{equation}

\subsection{Long Pulse Approximation for $\protect\xi (t)$}

In the long pulse case (in which $\alpha/\omega \ll 1$), simplified
expressions can be obtained for the integrals $I_i(t)$ in a way
similar that used in Ref.~\cite{Wang95}. For example, starting from
Eq.~(\ref{I1tvalu1AP}) one can expand the sine function in terms of
exponentials and then do the resulting integration by parts,
dropping terms that are of order $\alpha /\omega$ or higher, as
follows:
\begin{eqnarray}
I_{1}(t)&=&-\frac{\alpha}{\sqrt{\pi}}e^{i\omega _{c}t}\int_{-\infty
}^{t}e^{-\alpha ^{2}(t^{\prime}-\tau )^{2}}[e^{i (\omega-\omega_{c})
t^{\prime}+i\beta}- e^{-i
(\omega+\omega_{c}) t^{\prime}-i\beta} ] \label{I1expandedAP} \\
&\simeq&\frac{i\alpha}{\sqrt{\pi}}e^{-\alpha ^{2}(t-\tau
)^{2}}\left[ \frac{1}{%
\omega -\omega _{c}}e^{i\left( \omega t+\beta \right)
}+\frac{1}{\omega +\omega _{c}}e^{-i\left( \omega t+\beta \right)
}\right]. \label{I1approxAP}
\end{eqnarray}
In a similar way one may obtain the following approximate
expressions for $I_2(t)$ and $I_0(t)$:
\begin{eqnarray}
I_{2}(t) &\simeq &\frac{i\alpha}{\sqrt{\pi}}e^{-\alpha ^{2}(t-\tau
)^{2}}\left[ \frac{1}{%
\omega +\omega _{c}}e^{i\left( \omega t+\beta \right)
}+\frac{1}{\omega -\omega _{c}}e^{-i\left( \omega t+\beta \right)
}\right]
,\label{I2approxAP}\\
   I_{0}(t) &\simeq &\frac{-4i\alpha}{\sqrt{\pi
}}\frac{1}{\omega}e^{-\alpha ^{2}(t-\tau )^{2}}\cos \left( \omega
t+\beta \right). \label{I0approxAP}
\end{eqnarray}

Substituting these long pulse approximations for the integrals
$I_i(t)$ into Eq.~(\ref{ksait2AP}), the function $\xi (t)$ in the
long pulse limit is then given by Eq.~(\ref{longksi}).

%%%%%%%%%%%%%%%%%%%%%%%%%%%%%%%%%%%
%
%
% Acknowledgement Section
%
%
%%%%%%%%%%%%%%%%%%%%%%%%%%%%%%%%%%%
% If you have acknowledgments, this puts in the proper section head.
\begin{acknowledgments}
     We thank Ilya I. Fabrikant and Nikolai L. Manakov for critical readings of our manuscript.  L.Y.P.  acknowledges with appreciation many stimulating
discussions with Andrei Y. Istomin.
This work was supported in part by the Department of Energy,
Office of Science, Division of Chemical Sciences, Geosciences,
and Biosciences under Grant No. DE-FG02-96ER14646.
\end{acknowledgments}

%%%%%%%%%%%%%%%%%%%%%%%%%%%%%%%%%%%
%
%
% Bibliography Section
%
%
%%%%%%%%%%%%%%%%%%%%%%%%%%%%%%%%%%%

\bibliographystyle{AABBRV}

\bibliography{ACOMPAT,harvard}

\end{document}